\documentclass[sigconf]{acmart}

\AtBeginDocument{%
  }

\copyrightyear{2025}
\acmYear{2025}
\setcopyright{cc}
\setcctype{by-sa}
\acmConference[ISCA '25]{Proceedings of the 52nd Annual International Symposium on Computer Architecture}{June 21--25, 2025}{Tokyo, Japan}
\acmBooktitle{Proceedings of the 52nd Annual International Symposium on Computer Architecture (ISCA '25), June 21--25, 2025, Tokyo, Japan}\acmDOI{10.1145/3695053.3731096}
\acmISBN{979-8-4007-1261-6/2025/06}

\usepackage{makecell}
\usepackage{multirow}
\usepackage{siunitx}
\usepackage{subcaption}
\usepackage{enumitem}
\usepackage[font=footnotesize,labelfont=bf]{caption}
\usepackage{listings}
\usepackage{soul}
\usepackage{listings}
\usepackage{pifont}

\usepackage{acmart-taps}
\newcommand{\papertitle}{ATiM}
\newcommand{\cmark}{\textcolor{green!60!black}{\ding{51}}} 
\newcommand{\xmark}{\textcolor{red!75!black}{\ding{55}}} 

\definecolor{codegreen}{rgb}{0,0.6,0}
\definecolor{codegray}{rgb}{0.5,0.5,0.5}
\definecolor{codepurple}{rgb}{0.58,0,0.82}

\lstdefinestyle{customminted}{
    commentstyle=\color{codegreen},
    keywordstyle=\color{blue},
    numberstyle=\tiny\color{codegray},
    stringstyle=\color{codepurple},
    basicstyle=\selectfont\ttfamily\footnotesize,
    breaklines=true,
    captionpos=b,
    showspaces=false,
    showstringspaces=false,
    showtabs=false,
    tabsize=4,
    numbers=none
}

\begin{document}

\title{\papertitle: Autotuning Tensor Programs for Processing-in-DRAM}

\author{Yongwon Shin}
\authornote{Both authors contributed equally to this research.}
\orcid{0000-0002-0481-172X}
\email{ywshin@postech.ac.kr}
\affiliation{%
  \department{Graduate School of AI}
  \institution{POSTECH}
  \country{South Korea}
}
\author{Dookyung Kang}
\authornotemark[1]
\orcid{0009-0003-5643-5687}
\email{st00ne@snu.ac.kr}
\affiliation{%
  \department{Graduate School of Data Science}
  \institution{Seoul National University}
  \country{South Korea}
}
\author{Hyojin Sung}
\orcid{0000-0002-3036-6180}
\email{hyojin.sung@snu.ac.kr}
\affiliation{%
  \department{Graduate School of Data Science}
  \institution{Seoul National University}
  \country{South Korea}
}

\pagenumbering{gobble}
\newpage
\pagenumbering{arabic}
\setcounter{page}{1}
\begin{abstract}

Processing-in-DRAM (DRAM-PIM) has emerged as a promising technology for accelerating memory-intensive operations in modern applications, such as Large Language Models (LLMs). 
Despite its potential, current software stacks for DRAM-PIM face significant challenges, including reliance on hand-tuned libraries that hinder programmability, limited support for high-level abstractions, and the lack of systematic optimization frameworks. 
To address these limitations, we present \papertitle{}, a search-based optimizing tensor compiler for UPMEM.
Key features of \papertitle{} include: (1) automated searches of the joint search space for host and kernel tensor programs, (2) PIM-aware optimizations for efficiently handling boundary conditions, and (3) improved search algorithms for the expanded search space of UPMEM systems.
Our experimental results on UPMEM hardware demonstrate performance gains of up to 6.18$\times$ for various UPMEM benchmark kernels and 8.21$\times$ for GPT-J layers.
To the best of our knowledge, \papertitle{} is the first tensor compiler to provide fully automated, autotuning-integrated code generation support for a DRAM-PIM system. By bridging the gap between high-level tensor computation abstractions and low-level hardware-specific requirements, \papertitle{} establishes a foundation for advancing DRAM-PIM programmability and enabling streamlined optimization.

\end{abstract}

\maketitle

\section{Introduction}
\label{sec:introduction}

Modern applications are increasingly data-driven, leveraging vast amounts of data to optimize processes and improve efficiency. 
This has led to the inevitable shift of performance bottlenecks from compute unit efficiency to limited memory performance in computing system design.
Among various hardware and software-based approaches to continue performance scaling beyond the ``memory wall'', processing-in-DRAM~(DRAM-PIM) confronts the challenge by placing compute units near memory banks and computing directly on data in the memory instead of moving them. 
Recently, many hardware and memory vendors presented commercial and prototype DRAM-PIM products~\cite{devaux2019upmem,kim2021aquabolt,lee2022gddr6aim} to showcase their potential, providing up to 16$\times$ higher in-memory compute bandwidth without data transfer overheads. 
For example, DDR4-based UPMEM~\cite{devaux2019upmem} includes a RISC core called DPU (Data Processing Unit) per memory bank, allowing it to fully utilize the internal memory bandwidth and bank-level parallelism for accelerating memory-intensive workloads such as database systems, genomic analysis, and machine learning/deep learning inferences~\cite{chen2023uppipe,bernhardt2023pimdb,lim2023join,das2022dl,2023gomezlunaupmemtraining,gogineni2024swiftrl}.  

For DRAM-PIM to evolve into a mainstream architectural player,
a fully functioning, end-to-end software stack that provides user interfaces and code generation support for programmability and multi-level optimizations for performance is crucial.
However, current DRAM-PIM software stacks are still in an early and experimental phase, focusing on supporting a narrow set of target workloads with performance advantages over CPU/GPU. 
They provide a limited set of heavily hand-tuned libraries~\cite{kwon2022aimsoftware,lee2021hbmpim} for memory-intensive, PIM-friendly operations (e.g., GEMV) based on low-level programming model and compiler support~\cite{juan2022prim,jinfan2023simplepim}. 
Without high-level abstraction layers bridging the gap between such operations and different DRAM-PIM backends (e.g., C programming for UPMEM), substantial programming and engineering efforts must be repeated.

Recent research leveraged modern tensor-oriented intermediate representations (IRs) and compilers~\cite{ragankelley2013halide,vasilache2018tc,lattner2021mlir} to provide such abstractions.  
In tensor compilers, multi-dimensional arrays, i.e., tensors, are the base data type, and tensor IR instructions define operations repeated on tensor elements. 
Thus, tensor IRs can seamlessly represent in-memory computations as nested loops over tensors stored in the memory, facilitating lowering from high-level tensor operations to target-specific instructions.
Prior work leveraged tensor IRs to generate codes for different types of PIM hardware~\cite{khan2023cinm} or determine feasible mappings for tensor data to memory subarrays~\cite{drebes2020tccim}, focusing on improving the programmability of the PIM software stack.

Our key insight is that the true value of tensor compilation for DRAM-PIM lies in the unique optimization opportunities provided by domain-specific tensor abstractions: specifically, {\it enabling effective search-based optimization, i.e., autotuning}, and {\it facilitating aggressive transformations that exploit high-level semantic knowledge and hardware-specific constraints during tensor IR lowering.}

First, the autotuning framework can closely interface with the DRAM-PIM code generator to search the space of host and kernel tensor programs and optimize them simultaneously. 
Autotuning formulates optimizations as a search problem, exploring different optimized versions of a given tensor operation to identify the best-performing candidate through hardware measurements~\cite{chen2018tvm,chen1028autotvm,ansor2020zheng,feng2023tensorir}.
This approach can automatically generate codes optimized explicitly for target hardware, often outperforming expert-tuned libraries, thereby providing both programmability and performance benefits. 
Existing tensor compilers have shown that the autotuning framework can effectively optimize tensor kernel functions on CPUs and GPUs by sampling candidates with varying loop structures and optimization parameters such as loop tiling and unroll factors~\cite{ansor2020zheng,feng2023tensorir}. 
While CPU and GPU kernel functions can be optimized independently, UPMEM kernel optimization depends on host-code parameters as the host controls how to distribute tensors across memory banks.
This interdependency creates a larger and more complex search space, making the autotuning approach an even more compelling solution for optimal code generation for DRAM-PIM. 

In addition, tensor compilers can leverage target loop structures and DRAM-PIM hardware features for aggressive, correctness-guaranteed high-level optimizations. 
\sloppy{Specializing optimization passes for specific operations and hardware can enable optimizations that would be infeasible in low-level compilers without semantic context.} 
These PIM-aware optimizations complement autotuning and low-level approaches, offering unique performance benefits.

In this paper, we present \papertitle{}, an optimizing tensor compiler for modern DRAM-PIM.
\papertitle{} extends the autotuning framework~\cite{shao2022metaschedule} and tensor-level IR lowering and optimization support~\cite{feng2023tensorir} in the Apache TVM compiler~\cite{chen2018tvm} to provide fully automated search-based code generation for UPMEM~\cite{devaux2019upmem,fabrice2019upmem}. 
\papertitle{} expands the autotuner to sample candidates from a joint search space including host-to-DPU data distribution strategies and kernel loop-level optimization parameters. 
While TVM schedule primitives are originally designed for loop transformations in kernel loops, 
\papertitle{} repurposes them to simultaneously optimize UPMEM host and kernel operations. It also provides the lowering passes to translate these primitives to loop-based tensor IR programs.

Tightly integrating autotuning with code generation, in turn, enables \papertitle{} to handle the larger and more complex search space and reduce the evolutionary search overheads. It refines the evolutionary search mechanism and filters infeasible candidates early in the autotuning process by systematically leveraging UPMEM hardware constraints and parallel execution behaviors. 
In addition, \papertitle{} implements tensor-level code optimizations to aggressively reduce boundary check overheads, exploiting target algorithm and hardware knowledge. 
Experimental results show that \papertitle{}-compiled kernels outperformed hand-tuned UPMEM libraries~\cite{juan2022prim,jinfan2023simplepim} by up to 6.18$\times$ and 8.21$\times$ for various PIM benchmark kernels and GPT-J layers, respectively, through more efficient host and kernel code generation with optimized tiling factors and reduction strategies. 
To the best of our knowledge, \papertitle{} is the first effort to provide fully automated and autotuning-integrated code generation support for commercial DRAM-PIM within a streamlined tensor compilation flow.  
ATiM is publicly available at: \url{https://github.com/SNU-CODElab/atim.git}.

The main contributions of the paper are as follows:
\begin{itemize}
\item We develop a fully automated framework for search-based code generation for UPMEM systems. By reusing and extending TVM schedule primitives, \papertitle{} defines and explores the joint search space for host and kernel optimizations. It also provides code generation passes that translate autotuned schedule primitives to executables for UPMEM. 

\item We identify key performance bottlenecks in UPMEM systems and implement PIM-aware optimization passes at the tensor IR level. \papertitle{}'s boundary check elimination passes leverage high-level semantic knowledge and UPMEM's architectural features for significant performance gains. 

\item We enhance the TVM autotuning framework to handle the expanded and more complex search space for UPMEM. \papertitle{} refines the search mechanism to minimize sampling noises and biases early in the autotuning process,
improving the final result quality.

\end{itemize}
In the rest of the paper, we provide background information for DRAM-PIM architectures and software stacks in Section~\ref{sec:background}.
Section~\ref{sec:motivation} outlines motivations behind \papertitle{}, while Sections~\ref{sec:system_overview} and ~\ref{sec:design_and_implementation} detail the design and implementation of \papertitle{}. Section~\ref{sec:methodology} and~\ref{sec:evaluation} present the methodology and experimental results. Section~\ref{sec:discussion} discusses limitations and future directions, and Section~\ref{sec:related_work} reviews related work. Finally, Section~\ref{sec:conclusion} concludes the paper.
\section{Background}
\label{sec:background}

\subsection{Processing-in-DRAM~(DRAM-PIM)}

\begin{figure}[t]
    \includegraphics[width=\linewidth]{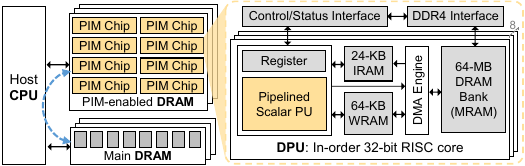}
    \vspace{-1.7\baselineskip}
    \caption{UPMEM architecture~\cite{devaux2019upmem}.}
    \vspace{-1.3\baselineskip}
    \label{fig:upmem_architecture} 
\end{figure}

Processing-in-memory technologies have a long history of influential research and industry proposals. While they share the ultimate goal of reducing data transfer overhead by computing where data is, they differ in processing unit (PU) design, memory types and structures, execution models, and software interfaces~\cite{mutlu2022modern,jeff2002diva,gokhale1995pim,nair2015memorycube,cho2021accel}.  
Among them, processing-in-DRAM or DRAM-PIM specifically targets extending DRAM architectures (including HBM) to incorporate compute units that can internally compute on data stored in the memory banks. 
However, it is only with recent advances in memory design and manufacturing technologies that several hardware and memory vendors have released their commercial and prototype DRAM-PIM products~\cite{lee2021hbmpim,lee2022gddr6aim,kim2022lpddr5pim,devaux2019upmem}.
While~\cite{lee2021hbmpim,lee2022gddr6aim,kim2022lpddr5pim} focus on closely integrating MAC acceleration logic with the memory array so that computations can actually happen on memory access paths, Devaux~\cite{devaux2019upmem} proposes putting a small core side by side with a data array per memory bank.  
This paper focuses on supporting the latter architecture with more general-purpose in-memory computing capabilities to showcase \papertitle{}'s tensor-level code generation and optimization capabilities.

\noindent
{\bf UPMEM DPU. }
The UPMEM architecture consists of a host CPU, standard DRAM main memory, and PIM-enabled DRAM, as shown in Fig.~\ref{fig:upmem_architecture}.
The PIM-enabled memory includes multiple chips, each with eight Data Processing Units (DPUs) capable of processing data. Since each DPU communicates directly with a single memory bank, UPMEM fully utilizes the internal bank bandwidth.
Each DPU features a 32-bit RISC-style pipelined in-order core with 24 threads (``tasklets'' for UPMEM), sharing a \SI{24}{KB} Instruction RAM (IRAM) for instructions and a \SI{64}{KB} Working RAM (WRAM) for operational data, alongside a Main RAM (MRAM) corresponding to the memory bank.
Data transfers between these components are managed by a DMA engine. 
In a UPMEM server, all data movement between the host CPU and DPUs must pass through the host CPU; even when data transfer between DPUs is required, it is routed via the host CPU. Because these transfers rely on limited memory channels between the host's main memory and MRAM, reducing data movement significantly impacts performance.

\noindent
{\bf DPU programming model. }
DPU programming is performed in an SPMD style using C, similar to OpenCL~\cite{stone2010opencl} or CUDA~\cite{nickolls2008cuda}, and the code is compiled via LLVM~\cite{lattner2004llvm}. Host code for DPU allocation, data management, and kernel execution can be written using the user interface provided by the UPMEM SDK.
Frameworks and libraries have been developed to simplify programming on UPMEM~\cite{juan2022prim, jinfan2023simplepim, khan2023cinm}.
PrIM~\cite{juan2022prim} proposes general programming guidelines for UPMEM and provides core kernel implementations as a library. It directly uses the C language and runtime functions provided by the UPMEM SDK.
SimplePIM~\cite{jinfan2023simplepim} proposes an efficient UPMEM programming method with a concise interface, inspired by the similarity between UPMEM and distributed computing—both distribute data and perform independent computations on each node (DPU). It simplifies the programming process by utilizing operations and communication primitives
(e.g., \texttt{map} and \texttt{reduce}).

\subsection{TensorIR}
\label{sec:tensorir}
TensorIR (TIR) is an intermediate representation (IR) in the TVM compiler stack~\cite{chen2018tvm} designed specifically to generate and optimize high-performance code for tensor computations~\cite{feng2023tensorir}.
TensorIR's main idea is to separate high-level computational definitions from the implementation details. TensorIR provides users with Python-based programming interfaces to define abstract computational tasks and let separate ``schedule primitives'' determine how the computation is optimized and compiled to low-level codes. 

\begin{figure}[t]
    \includegraphics[width=\linewidth]{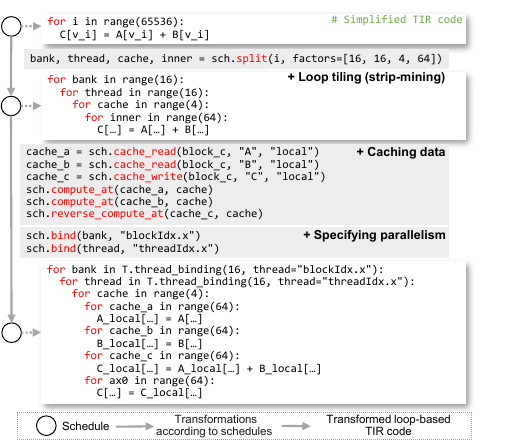}
    \vspace{-1.7\baselineskip}
    \caption{TIR transformations according to schedule primitives.}
    \vspace{-0.7\baselineskip}
    \label{fig:schedule_primitive} 
\end{figure}

\noindent
{\bf Schedule primitives.}
Inspired by Halide~\cite{ragankelley2013halide}, TVM schedule primitives can specify flexible loop transformations without writing complex codes.
Fig.~\ref{fig:schedule_primitive} demonstrates how different schedule primitives shape loop structures. 
The \texttt{split} and \texttt{reorder} primitives handle loop tiling and reordering. 
While \texttt{cache\_read} and \texttt{cache\_write} mark data to be cached and written back, \texttt{compute\_at} and \texttt{reverse\_ \linebreak compute\_at} determine its caching locations within loops.
To enable parallel execution, the \texttt{bind} primitive assigns loops to threads. 
In Fig.~\ref{fig:schedule_primitive}, the outermost loop is bound to \texttt{blockIdx.x} (representing a thread block, as in GPU programming), and the immediate inner loop is bound to \texttt{threadIdx.x} (individual threads).
These schedule primitives can be combined to generate codes with multiple transformations applied simultaneously.

\noindent
{\bf TIR lowering process. }
Although there is no official distinction, schedule primitives are lowered to TIR with explicit loops, i.e., loop-based TIR, in the process of lowering TIR to target hardware. While high-level TIR representations include metadata such as block and tensor information along with schedule primitives, they are translated into an imperative language code format in loop-based TIR to facilitate low-level IR generation such as LLVM IR. A loop-based TIR program is further lowered to separate TIR programs for host and DPU kernels, which are provided as input for the UPMEM backend.

\noindent
{\bf Autotuning support.}
Autotuning is a search-based optimization technique that explores the code space for optimal performance based on profiled runs. TVM provides autotuning support at the TensorIR level~\cite{shao2022metaschedule,feng2023tensorir}. 
TVM generates autotuning candidates in the search space by applying different schedule primitives and assigning random values to parameters. It then repeatedly measures them on hardware to find the optimal version. The searches are guided by a performance prediction model. 

\section{Motivating Observations}
\label{sec:motivation}

In this section, we present key observations about current software stacks and hardware behaviors of UPMEM that motivate and guide the design of \papertitle{}.

\noindent
\textit{\textbf{Current software stacks for UPMEM provide only low-level programming models with limited high-level abstractions, requiring substantial development and tuning effort.}}
Offloading computations to UPMEM involves programming tasks similar to GPU offloading with OpenCL or CUDA: 
writing host code to distribute data, launch kernels, and post-process results as well as PIM kernel functions for each DPU. 
DPU programming is even more demanding and labor-intensive than GPU programming because UPMEM lacks hardware runtime features to provide flexible resource mapping and address translation, e.g., hardware scheduler or special base address registers, which forces developers to exactly address host and kernel data.  
Library-based approaches~\cite{juan2022prim} offer reusable building blocks to simplify programming, but host and kernel codes are still explicitly written and tuned for a specific workload, often in several hundreds of lines.

High-level abstractions such as~\cite{jinfan2023simplepim,khan2023cinm} can provide more extensible and general-purpose code generation support, but current solutions only provide a limited set of abstractions or fail to fully automate the process. 
For example, SimplePIM~\cite{jinfan2023simplepim} only provides one-dimensional tensor abstractions,
requiring extra effort to handle higher-dimensional data.
While CINM~\cite{khan2023cinm} supports tensor operations such as GEMV, GEMM, and histogram with multi-level IRs to automate code generation, it still requires programmer intervention for tasks like aggregating DPU data on the host. 
Moreover, no existing solutions provide optimization passes that systematically transform host and kernel codes for DRAM-PIM efficiency, leaving their claims for potential tensor-level optimizations unverified.
To address these gaps, we introduce \papertitle{}, which offers flexible code generation for both host and kernel codes, enabling search-based optimizations with manual tuning.

\begin{figure}[!t]
    \centering    \includegraphics[width=\linewidth]{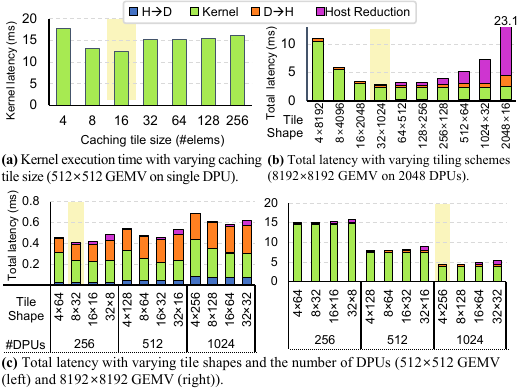}
    \vspace{-1.7\baselineskip}
     \caption{Performance impact of caching tile sizes, tiling schemes, and the number of DPUs.} 
    \label{fig:motiv_space}
    \vspace{-1.1\baselineskip}
\end{figure}

\begin{figure}[!t]
    \centering
    \includegraphics[width=\linewidth]{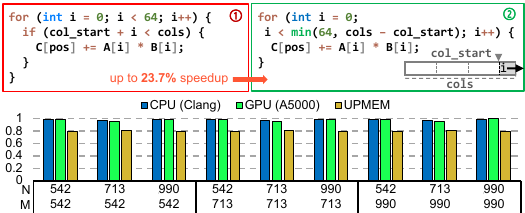}
    \vspace{-1.7\baselineskip}
    \caption{Boundary checks' impact on GEMV kernel (M $\times$ N)  execution time.} 
    \label{fig:motivbranch}
    \vspace{-1.2\baselineskip}
\end{figure}

\noindent 
\textit{\textbf{Intra-DPU and inter-DPU optimizations have a vast search space of closely correlated parameters with significant performance impact.}}
Preliminary experiments (Fig.~\ref{fig:motiv_space}) reveal that DPU performance is significantly impacted by {\it inter-DPU parallelism}, such as how data and computations are distributed across how many DPUs, as well as {\it intra-DPU optimizations} including loop parallelization, tiling, and caching strategies within each DPU.
At the inter-DPU level, how tensor data is partitioned and distributed across DPUs (``tile size'') determines kernel execution and host reduction time, with optimal tile sizes minimizing both (Fig.~\ref{fig:motiv_space}(b)). 
The performance impact of tile sizes also depends on the number of DPUs used relative to the original tensor's size and shape; for smaller tensors (e.g., 512$\times$512), fewer DPUs than the maximum available can result in better performance (Fig.~\ref{fig:motiv_space}(c)).
Once data is distributed, intra-DPU optimization parameters, such as tasklet parallelism, loop tiling and unroll factors, and WRAM size, are critical for DPU kernel performance (Fig.~\ref{fig:motiv_space}(a)). 
Optimized kernel execution time, in turn, affects inter-DPU performance trends, creating a closely correlated optimization space between the two levels. 
Previous work showed the benefits of heuristic-based approaches for optimizing inter-DPU data distribution~\cite{jinfan2023simplepim}. 
In contrast, \papertitle{} focuses on systematic autotuning, expanding the optimization space with PIM-specific parameters from both intra-DPU and inter-DPU optimizations to identify the most efficient configurations. 

\noindent
\textit{\textbf{UPMEM compute units can suffer from underutilization due to unoptimized branches. }}
Since DPU kernels for tensor computations iteratively fetch tensor data from MRAM to WRAM at high internal DRAM bandwidth, maximizing DPU core utilization is crucial to maximize the overall compute efficiency. However, prior work observed that simple in-order DPU cores without latency-hiding hardware make the system strongly compute-bound and highly susceptible to severe underutilization from frequent branch instructions~\cite{upimulator}. 
While uPIMulator~\cite{upimulator} explored architectural improvements to mitigate branch penalties, our focus is on code-level optimizations to eliminate branches. 
As shown in Fig.~\ref{fig:motivbranch}, our preliminary experiments show that eliminating a redundant boundary check can provide up to a 23.7\% speedup for GEMV operations. The caveat is that such optimizations are not always applied automatically by low-level compilers like LLVM, which lack high-level loop and dependency information.
Furthermore, this performance impact is particularly prominent for PIM architectures, where limited compute resources amplify the benefits of software-level branch optimizations. As shown in Fig.~\mbox{\ref{fig:motivbranch}}, CPUs and GPUs see minimal performance gains, likely due to advanced hardware branch handling and latency-hiding mechanisms, while UPMEM achieves an average 20\% runtime reduction across different input sizes. 
This highlights the need for aggressive optimizations at the tensor level to specifically optimize for DPU core utilization.

To address these challenges, \papertitle{} provides an extensible, optimizing tensor compiler for UPMEM. As shown in Table~\ref{table:upmem_support_comparison}, \papertitle{} supports fully automated code generation for a broad range of operations with diverse tensor shapes and dimensions, while enabling autotuning in the intra-DPU and inter-DPU optimization space -- capabilities that are not comprehensively supported by prior work.

\section{System Overview}
\label{sec:system_overview}

\begin{scriptsize}
\begin{table}[!t]
    \centering
    \caption{Features supported by UPMEM and prior work.}
    \vspace{-1.2\baselineskip}
    \label{table:upmem_support_comparison}
    \renewcommand{\arraystretch}{1.1}
     \begin{tabular}{|c|c|c|c|
      c|
    }
    \hline
    \bfseries & \bfseries PrIM & \bfseries SimplePIM & \bfseries CINM & \bfseries \papertitle{} (ours) \\
    \hline
    \hline
    \bfseries High-level programming abstractions & \xmark & \cmark & \cmark & \bfseries \cmark \\
    \hline
    \bfseries High-dimensional support & \xmark & \xmark & \cmark & \bfseries \cmark \\
    \hline
    \bfseries Inter-DPU optimization & \xmark & \xmark & \cmark & \bfseries \cmark \\
    \hline
    \bfseries Intra-DPU optimization & \cmark & \xmark & \cmark & \bfseries \cmark \\
    \hline
    \bfseries PIM-aware optimization & \cmark & \cmark & -- & \bfseries \cmark \\
    \hline
    \bfseries Autotuning support & \xmark & \xmark & \xmark & \bfseries \cmark \\
    \hline
    \end{tabular}
    \vspace{-2.0\baselineskip}
\end{table}
\end{scriptsize}

\begin{figure*}[!ht]
    \includegraphics[width=\textwidth]{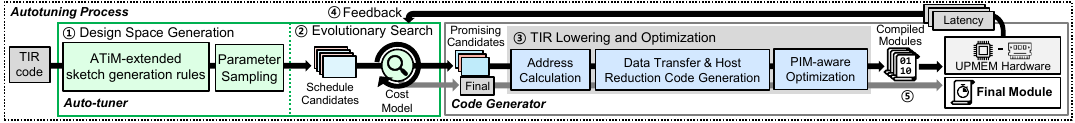}
    \vspace{-1.9\baselineskip}
    \caption{The overall system overview of \papertitle{}.}
    \label{fig:overview}
    \vspace{-1.0\baselineskip}
\end{figure*}

Fig.~\ref{fig:overview} illustrates the process of generating optimized UPMEM host and kernel codes from high-level operations through an autotuning framework and PIM-aware tensor-level transformations.
The autotuner (green box) and the code generator (gray box) are technically separate components, since the code generator can also take arbitrary tensor programs as input, including hand-written schedule primitives. 
However, \papertitle{} assumes the autotuner always initiates the code generation process (\textcircled{1} to \textcircled{4}), except for the final compilation step when the code generator directly compiles host and kernel binaries using autotuning results (\textcircled{5}). 

When given a target operation to offload, the autotuner generates schedule primitive sequences using the sketch generation rules and sampled optimization parameters to populate the space of possible host and kernel code implementations. 
Once these rules generate schedule primitives to determine code structures, tunable parameter values are sampled within loop bounds. 
Then, \papertitle{} performs an evolutionary search guided by a cost model to identify promising candidates to compile and evaluate on hardware. 
\papertitle{} extends the existing TIR lowering passes in TVM to translate schedule primitives to represent correct tensor programs for UPMEM host and kernel operations. \papertitle{} implements address calculation logic for kernel memory accesses to tiled data per bank, and added PIM-specific TIR lowering passes for kernel and host reduction loops and host data transfer function calls. 
\papertitle{} also applies PIM-aware optimizations that leverage high-level semantic knowledge and PIM hardware constraints. 
The resulting host and kernel programs are compiled by the TVM backend using the LLVM compiler and TVM runtime, and evaluated by the autotuner on UPMEM hardware to collect performance results and train the cost model to guide the search.

\begin{scriptsize}
\begin{table*}[!t]
    \centering
    \caption{Tunable host and kernel operations with corresponding schedule primitives.}
    \vspace{-1.3\baselineskip}
    \label{table:schedule}
    \renewcommand{\arraystretch}{1.1}
    \begin{tabular}{|p{0.12\linewidth}|p{0.14\linewidth}|p{0.13\linewidth}|p{0.1\linewidth}|p{0.40\linewidth}|}
    \hline
    \bfseries & \bfseries Schedule primitives & \bfseries Tunable parameters & \bfseries Codegen target & \bfseries Examples \\
    \hline
    \hline
    \multirow{3}{*}{\parbox[t]{\linewidth}{Host-to-DPU \\ data distribution}} & \texttt{split} & tiling factor & host & \parbox[t]{\linewidth}{\ttfamily x\_dpu, x\_in\_dpu = sch.split(x, factors=[XB, None]) \\ y\_dpu, y\_in\_dpu = sch.split(y, factors=[YB, None])} \\
    \cline{2-5}
    & \texttt{reorder} & - & host & \ttfamily sch.reorder(x\_dpu, y\_dpu, x\_in\_dpu, y\_in\_dpu) \\
    \cline{2-5}
    & \texttt{bind} & - & host & \parbox[t]{\linewidth}{\ttfamily sch.bind(x\_dpu, "blockIdx.x") \\ sch.bind(x\_dpu, "blockIdx.y")} \\
    \hline
    Reduction strategy & \texttt{rfactor} & - & host, kernel & \ttfamily block\_dpu = sch.rfactor(x\_dpu, factor\_axis=0) \\
    \hline
    \multirow{3}{*}{Multi-level tiling} & \texttt{split} & tiling factor & kernel & \parbox[t]{\linewidth}{\ttfamily thread\_dpu, y\_cache, y\_in = sch.split(y\_in\_dpu, factors=[...]) \\ x\_cache, x\_in = sch.split(x\_in\_dpu, factors=[...])} \\
    \cline{2-5}
    & \texttt{reorder} & - & kernel & \ttfamily sch.reorder(thread\_dpu, y\_cache, y\_in, x\_cache, x\_in) \\
    \cline{2-5}
    & \texttt{bind} & - & kernel & \ttfamily sch.bind(thread\_dpu, "threadIdx.x") \\
    \hline
    \multirow{2}{*}{Intra-DPU caching} & \texttt{cache\_read/write} & - & kernel & \parbox[t]{\linewidth}{\ttfamily cache\_a = sch.cache\_read(block\_dpu, 0, "local") \\ cache\_b = sch.cache\_read(block\_dpu, 1, "local") \\ cache\_c = sch.cache\_write(block\_dpu, 0, "local")} \\
    \cline{2-5}
    & \texttt{(reverse\_)compute\_at} & caching location & kernel & \parbox[t]{\linewidth}{\ttfamily sch.compute\_at(cache\_a, x\_cache) \\ sch.compute\_at(cache\_b, x\_cache) \\ sch.reverse\_compute\_at(cache\_c, y\_cache)} \\
    \hline
    \multirow{2}{*}{Post-processing} & \texttt{split} & tiling factor & host & \ttfamily thread, \_ = sch.split(y, factors=[...]) \\
    \cline{2-5}
    & \texttt{parallel} & - & host & \ttfamily sch.parallel(thread) \\
    \hline
    \end{tabular}
    \vspace{-1.2\baselineskip}
\end{table*}
\end{scriptsize}

\section{\papertitle{}: Design and Implementation}
\label{sec:design_and_implementation}

\subsection{Tunable Host and Kernel Operations}
Offloading tensor computations to UPMEM involves a sequence of host and kernel operations that, at a high level, follow a standardized process. 
We first outline these actions at an abstract, algorithmic level without specifying implementation details, so that we can separate out detailed ``schedules'' and optimize them through targeted searches~\cite{feng2023tensorir}. 

Host code determines how to tile input and output tensor data across DPUs and aggregate results. This is a global optimization problem with intractable complexity, with conflicting trade-offs between intra- and inter-DPU parallelism and communication/synchronization overheads. \papertitle{} addresses this by identifying key tunable parameters and applying search-based optimizations to generate efficient host code. Tiling strategies, parallelism levels, and host-to-DPU data transfer granularity impact both kernel performance and host-DPU communication cost, while thread-level parallelism can accelerate the post-processing on the host. 

On the other hand, autotuning DPU kernel functions focuses mainly on optimizing loop structures, much like GPU kernel autotuning. Loop structures define which loops are parallelized (spatial loops) and which cache intermediate results (reduction loops). Unrolling and multi-level tiling factors affect tasklet-level parallelism and WRAM locality, while MRAM-WRAM caching locations and sizes decide intra-DPU data transfer overheads. Reduction strategies affect both host and kernel code operations, depending on whether results are partially reduced on DPUs first then aggregated on the host, or sent directly to the host. 

In the following sections, we describe how we extend the existing tensor compiler to support tunable code structures and parameters for UPMEM, search for optimal schedules, and lower them to host and kernel implementations.

\subsection{Search-based Code Generation}
\label{sec:search_based_code_generation}

\subsubsection{Schedule Candidate Generation}
\label{subsec:schedule_candidate_generation}
\begin{figure*}[!t]
    \includegraphics[width=\linewidth]{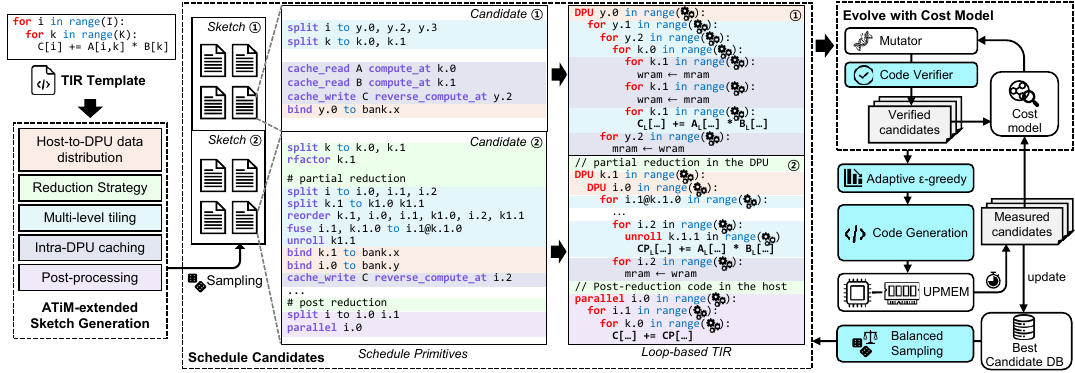}
    \vspace{-2.0\baselineskip}
    \caption{Autotuning-driven code generation process in \papertitle{}.}
    \label{fig:autotuning_process}
    \vspace{-1.2\baselineskip}
\end{figure*}

~~~\\
\papertitle{} extends the sketch generation rules~\cite{feng2023tensorir} in the TVM autotuning framework to generate schedule primitive samples that implement essential host and kernel operations with varying code structures and optimization parameters (Fig.~\ref{fig:autotuning_process}). 
The core idea behind \papertitle{}'s schedule candidate generation is {\it repurposing TVM schedule primitives for UPMEM host and kernel code generation.}
While preserving much of the original semantics of these schedule primitives, \papertitle{} leverages them specifically to define and populate the search space for UPMEM optimizations, as shown in Table~\ref{table:schedule}.

\noindent
{\bf Host-to-DPU data distribution. }
\papertitle{} tiles tensor data and maps them across DPUs by using \texttt{split} and \texttt{reorder} schedule primitive and annotating tiled loop variables to bind them with DPUs.  
Unlike prior approaches~\cite{gu2020ipim} that introduced fixed data distribution strategies with separate schedule primitives, \papertitle{} enables flexible data mapping and distribution by leveraging existing loop schedule primitives.
\texttt{split} and \texttt{reorder} schedule primitives are used to define the loop tiling pattern, i.e., tile sizes in each dimension and the ordering between them.
In the example, \texttt{split} primitives tile the loop to iterate over \texttt{XB}$\times$\texttt{YB} tiles, and \texttt{reorder} primitive specifies the row-major traversal order. 
Schedule candidates also perform ``DPU binding'' through \texttt{bind} primitive to specify which loop level corresponds to inter-DPU parallelism, i.e., determine tiles per DPU. The bank index uses linearized values of the loop variables with the annotation. In the example, the inner tiles are distributed to \texttt{XB}$\times$\texttt{YB} DPUs by binding the loop variables \texttt{x\_dpu} and \texttt{y\_dpu}.

\noindent
{\bf Reduction strategy. }
When multidimensional tiles are mapped to DPUs, computation results can be partially reduced in one dimension, and intermediate results can be aggregated on the host. \papertitle{} uses \texttt{rfactor} schedule primitives to represent such parallel reduction strategies on UPMEM. 

\noindent
{\bf Multi-level tiling.  }
\papertitle{} generates a sequence of \texttt{split}, \texttt{reorder}, and \texttt{bind} primitives to define how to execute the kernel loop at each DPU. For example, \texttt{split} and \texttt{reorder} perform multi-level tiling by creating an outermost loop for multi-tasklet execution (\texttt{thread\_dpu}), additional loop levels to cache data (\texttt{x\_cache} and \texttt{y\_cache}), and innermost loops (\texttt{y\_in} and \texttt{x\_in}). 
Then, \texttt{bind} primitives associate the outermost loop iterations with parallel thread IDs to exploit intra-DPU thread parallelism (``tasklet binding''). 

\begin{figure}[!t]
    \includegraphics[width=\linewidth]{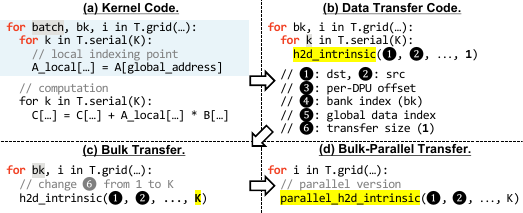}
    \vspace{-1.8\baselineskip}
    \caption{Data transfer code generation and optimization.}
    \label{fig:data_transfer_codegen_and_opt}
    \vspace{-1.6\baselineskip}
\end{figure}

\noindent
{\bf Intra-DPU caching. }
DPU kernels must explicitly load and store data in WRAM before and after computation. 
MRAM-WRAM caching locations and granularities (i.e., caching data sizes) significantly impact data locality and transfer overheads;
per-thread caching may be too coarse-grained for WRAM, limiting parallelism, while caching inside the innermost tiling loop can trigger excessive MRAM-DRAM data transfers. 
\papertitle{} allows the autotuner to search for optimal caching locations and granularities by using \texttt{cache\_read/write} and \texttt{(reverse)\_compute\_at} schedule primitives.
\sloppy{\texttt{cache\_read/write} primitives define WRAM tiles to load or store, and \texttt{(reverse)\_compute\_at} binds these caching tiles with loops to specify exact caching locations.} 

\noindent
{\bf Post-processing. }
After DPU kernel execution, the host code aggregates computation results from DPUs to produce the final output. This loop can incur significant overhead if executed sequentially for a larger number of DPU results. \papertitle{} enables tiling and parallel optimizations for this host post-processing loop by using \texttt{split} and \texttt{parallel} schedule primitives. 

\subsubsection{TIR Lowering}
\label{subsec:tir_lowering}
~~~\\
The code generator for UPMEM lowers schedule primitive sequences selected from the evolutionary search to loop-based TIR programs.
While \papertitle{} reuses the current TIR lowering passes to construct nested loops, apply loop transformations such as unrolling and tiling, and generate parallel DPU kernels and tasklets per DPU, it performs the following \papertitle{}-specific passes to generate per-DPU address calculation (kernel), data transfer function calls (host), and partial and final reduction loops (host and kernel). Final TIR programs are compiled to host and kernel binaries by the Clang/LLVM compiler in the UPMEM SDK and TVM runtime.  

\noindent
{\bf Address calculation.}
Address calculation also leverages schedule primitives to introduce a local DPU scope in compile time. 
Without runtime translation support such as base address registers, calculating a per-DPU offset from a global address is tedious and can incur high overheads for complicated tiling patterns, e.g., high-dimensional tiles or tiles with strides. 
By allocating caching tiles for WRAM using \texttt{cache\_read/write} and specifying their locations using \texttt{compute\_at}, \papertitle{} allows DPU-local data on WRAM to be accessed simply by adding outer-loop indices multiplied by the caching tile size (base address) and the inner-loop indices (per-DPU offset). Since tiled data in MRAM maintains its host address, \papertitle{} can quickly establish the mapping from the global tensor address to the per-DPU local address. 

\noindent
{\bf Data transfer code generation.}
ATiM reuses the global-to-local address mapping established in kernel codes to generate data transfers. 
As shown in Fig.~\ref{fig:data_transfer_codegen_and_opt}, starting from the kernel TIR code generated for UPMEM (a), we adapt the loop nest to perform iterative but non-redundant data transfers by removing the outer loop \texttt{batch}, keeping only the loops outside the local indexing point (loops \texttt{bk}, \texttt{i}, and \texttt{k}), and putting a \papertitle{} data transfer intrinsic at the WRAM caching location (b).
For host-to-device transfers, this intrinsic takes the global data index as a source and the DPU index and per-DPU offset as a destination. Device-to-host transfer codes can be generated similarly, with the transfer direction reversed. 

While the code in Fig.~\ref{fig:data_transfer_codegen_and_opt}(b) works correctly, its performance can be improved by allowing parallel and bulk transfers. 
We implement the following two optimization passes to generate more efficient data transfer codes for UPMEM (as shown in Fig.~\ref{fig:data_transfer_codegen_and_opt}(c) and (d)). 

\begin{itemize}[leftmargin=*,parsep=0in,topsep=0.05in]
    \item \textit{Bulk transfer}: \papertitle{} coalesces individual data transfers to a single transfer of a contiguous data chunk. This optimization pass visits loops with data transfer intrinsics, checks if data accesses are contiguous within a loop nest, and generates ``vectorized'' transfers by unrolling the loops. This process is repeated until further unrolling is impossible. As shown in Fig.~\ref{fig:data_transfer_codegen_and_opt}(c), loop \texttt{k} is unrolled to increase the data transfer size from 1 to \texttt{K}, \texttt{k}'s loop extent.
    
   \item \textit{Bank-parallel transfer}: For UPMEM with parallel data transfer support, \papertitle{} provides parallel data transfer intrinsics. Parallel transfers of data banks (DPU) in a rank are supported by \texttt{dpu\_prepare\_xfer} and \texttt{dpu\_push\_xfer} functions, to which parallel data transfer intrinsics in Fig.~\ref{fig:data_transfer_codegen_and_opt}(d) are later lowered. 
\end{itemize}

\noindent
{\bf Reduction code generation.} 
The \texttt{rfactor} primitive is originally for securing additional parallelism in the reduction dimension in kernel codes. \papertitle{} code generator translates it to hierarchical reduction with parallel partial reduction loops across DPUs and a final reduction loop on the host. 
The resulting loops reduce the amount of input data fetched by each DPU while generating another reduction loop over intermediate output vectors. The cost of executing the additional final reduction loop on the host CPU is quickly offset by reduced data transfers between the host and DPUs. The final reduction loop can be further tiled and executed in parallel on multi-threaded CPUs. 

\begin{figure*}[!ht]
    \includegraphics[width=\linewidth]{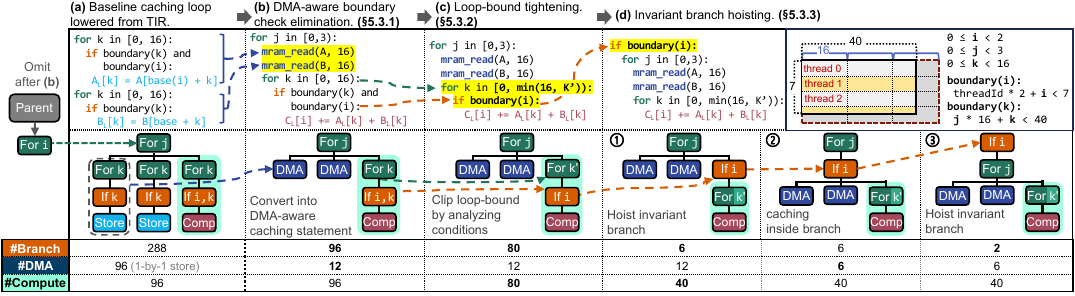}
    \vspace{-1.7\baselineskip}
    \caption{Example of PIM-aware optimizations applied to a 7$\times$40 GEMV kernel (single DPU, 4 threads with 4-byte data type). The kernel processes two rows at a time to meet the 8-byte alignment for the output and caches 16 elements per iteration, forming a 2$\times$16 tiling pattern, introducing boundary conditions along the row and column axes. The table shows the number of branches, DMA transfers, and innermost loop executions after each optimization step. }
    \label{fig:tiropt}
    \vspace{-1.0\baselineskip}
\end{figure*}

\subsubsection{Balanced Evolutionary Search}
\label{subsection:balanced_evolutionary_search
}
~~~\\
\papertitle{} augments the evolutionary search mechanism in the current auto-tuner to explore the expanded search space more effectively, as shown in Fig.~\ref{fig:autotuning_process} (right).
While autotuning for CPU or GPU focuses solely on the search space from computational subgraphs, \papertitle{}'s search space additionally includes tunable host operations, especially host-to-DPU data distribution, which determines inter-DPU parallelism. As a result, the evolutionary search for \papertitle{} must sample schedule candidates to simultaneously optimize inter-DPU parallelism across DPUs and intra-DPU (tasklet) parallelism per DPU kernel. 

We observed that this more extensive and complex search space biases the search toward inter-DPU parallelism over intra-DPU parallelism, with orders of magnitude more DPUs than per-DPU tasklets in typical UPMEM systems. 
Early in the autotuning process, this bias leads to overly favoring schedule primitives with \texttt{rfactor} that exploit inter-DPU parallelism for hierarchical reduction while prematurely dropping out candidates without \texttt{rfactor} from the search space, resulting in suboptimal performance. 

To address this \texttt{rfactor} primitive bias, \papertitle{} employs two techniques: \textit{balanced sampling} and an \textit{adaptive epsilon-greedy strategy}. The balanced sampler ensures equal representation of candidates with and without \texttt{rfactor} during early autotuning iterations. Specifically, it selects an equal proportion of top-K candidates for both design spaces (\textcircled{1} and \textcircled{2} in Fig.~\ref{fig:autotuning_process}) from the best candidate database.
This technique promotes exploration and preserves non-\texttt{rfactor} candidates but is only applied during the first 40\% of total trials to balance exploration and exploitation. After this point, top-K candidates are extracted without distinguishing between design spaces to accelerate convergence. Candidates directly sampled from the design spaces follow uniform distribution across design spaces, hence they do not require the balanced sampler.

The adaptive epsilon-greedy strategy increases the epsilon value during early autotuning phases, allowing diverse schedule candidates to be added to the best candidate database. However, maintaining a high epsilon value indefinitely can hinder convergence. Therefore, \papertitle{} starts with an epsilon value of 0.5 and linearly decreases it to the default value of 0.05 after the first 40\% of trials.

\subsubsection{Code verifier for UPMEM}
~~~~\\
\papertitle{} includes a schedule verifier to filter out candidates that violate UPMEM constraints during evolutionary search. Compared to CPUs and GPUs, UPMEM imposes much stricter hardware and programming constraints. Unlike GPUs, which manage thread blocks via hardware schedulers, UPMEM requires explicit management of up to 2560 DPUs and 24 tasklets per DPU, with a \SI{64}{KB} WRAM limit for caching. This often leads to invalid candidates that waste resources and degrade autotuning efficiency during an evolutionary search if unchecked. By eliminating such candidates early, the verifier reduces overhead and ensures high-quality results.

\subsection{PIM-aware Optimizations}
\label{sec:kernel_code_optimization}

\papertitle{} introduces tensor-level optimizations that leverage target loop structures and PIM hardware features to streamline control flows and improve PIM utilization. 
During kernel code generation, the TIR lowering passes generate loop-based TIR for tiled tensors and insert conditional statements to ensure that DPU tasklets access only valid memory within tensor boundaries.
These boundary checks do not affect interior tiles, 
but they prevent memory overflows or unnecessary computations for tiles at dimension boundaries whose extents may not align with the tile size. 

While boundary checks do not significantly impact performance on traditional processors thanks to latency-hiding techniques (e.g., branch predictors, ILP/TLP support), 
they can cause severe front-end stalls for simpler in-order DPU cores, which cannot mitigate control hazards without branch predictors and reorder buffers~\cite{upimulator}. 
Strict area and power constraints on DRAM chips would not easily allow such complicated logic within a DPU. As a result, boundary checks can have a lasting, damaging impact on DPU kernel performance. 

Software-level optimizations to eliminate or reduce boundary checks can be essential to minimize branch overheads in this case. 
While low-level compilers like LLVM can simplify control flow (e.g., merging, eliminating, or hoisting branches), most boundary checks remain unsafe to optimize at this stage, except for loop invariant code motion (LICM) partially applied to boundary equations, without high-level semantic knowledge. 
On the other hand, algorithm-level optimizations can enable more aggressive code rewriting by specifically targeting boundary checks. DietCode~\cite{dietcode} proposed two optimization techniques: (1) padding data locally at tile boundary or globally per tensor data to ensure out-of-bound accesses are safe, thereby removing the need for boundary checks, and (2) partitioning the loop body into slow boundary and fast non-boundary sections~\cite{shen2021nimble}. 
While TVM implements loop partitioning, it provides limited benefits for DPU kernels, as it is effective only for large tensor shapes.
Data padding can eliminate boundary checks but significantly increase storage and transfer overheads, particularly with global padding.

\papertitle{} adopts the padding techniques~\cite{dietcode} and introduces three TIR-level transformation passes that collaboratively reduce boundary checks and optimize memory accesses guarded by them using DPU functions. 
These passes target loop-based TIR kernel codes with affine access patterns and static tensor shapes, generated in the TIR lowering process, which are main optimization targets for tensor compilers and DRAM-PIM.
These loops are not part of the autotuning targets, as the current TVM autotuner performs loop tiling in a way that prevents misaligned tiles.
Since the TIR semantics for \texttt{compute\_at} and \texttt{reverse\_compute\_at} schedule primitives enforce that all consumer operations of a loop are under the given loop's constraints~\cite{feng2023tensorir},
we can assume that target loops have no other statements outside of the boundary condition.
This specific loop structure enables aggressive loop transformation and branch elimination, which low-level compilers cannot reliably validate.

\subsubsection{DMA-aware boundary check elimination}
\label{subsection:optimization_caching}
\hfill\\\
This optimization pass eliminates redundant boundary checks for load and store instructions for WRAM. These instructions fetch operands to WRAM from MRAM before computation and write them back to the memory bank afterward. Once boundary checks are removed, the enclosing loops become vectorizable, so we can replace each loop with a DMA instruction between WRAM and the memory bank.
DMA instructions can significantly speed up data transfers for contiguous data, but they can only be used without condition statements.
As shown in Fig.~\ref{fig:tiropt}(b), this pass can replace a loop with contiguous store instructions with a DMA instruction after eliminating the boundary check.

We can safely remove these boundary checks because 
(1) memory regions on DPU MRAM are locally padded, i.e., allocated in multiples of tile sizes, preventing out-of-bound accesses between MRAM and WRAM from corrupting meaningful data and 
(2) the same boundary checks are repeated later, and as long as we keep the ones guarding the computation in the kernel and readout in the host, we ensure that host data remains protected.
While this may slightly increase data transfer traffic outside the boundary, the performance gains far outweigh its minimal impact. 

\subsubsection{Loop-bound tightening}
\label{subsection:loopextent}
\hfill\\\
When a boundary check condition right after the loop header is a linear inequality and there are no other instructions in the loop body, which are assumed results of the current TIR lowering mechanism, this condition can intersect with the loop's upper bound condition by solving the two linear inequalities.
As illustrated in Fig.~\ref{fig:tiropt}(c), this process tightens the upper bound of the loop variable \texttt{k} after being merged with the following boundary condition (\texttt{if boundary(k) and boundary(i)}). 
Foregoing ``dead'' iterations known to fail the boundary check in compile time can significantly reduce the loop execution time (from 96 to 80 iterations). The computation complexity of the loop-bound evaluation may increase, which can later be optimized through strength reduction. 
General-purpose compilers lack equivalent loop optimization passes, while previous research on polyhedral compilation exploited user-provided assumptions~\cite{grosser2012polly} or directives~\cite{clangdirectives} to enable broader and more advanced loop optimizations. In contrast, \papertitle{} can safely apply loop-bound tightening without requiring additional safety checks or user input by focusing on lowered IRs with structural properties guaranteed by the IR lowering process, rather than arbitrary user-written code. 
Once the optimization is applied, only a boundary check with a loop invariant variable (\texttt{i}) remains, which can be further optimized by invariant branch hoisting in Section~\ref{subsection:hoist}. 
\subsubsection{Invariant branch hoisting}
\label{subsection:hoist}
\hfill\\\
Invariant branch hoisting moves boundary checks with loop-invariant conditions, which remain after loop-bound tightening, outside the loop \texttt{k} (step \textcircled{1}), as shown in Fig.~\ref{fig:tiropt}(d).
Invariant branches can also be hoisted out of the loop later by a low-level compiler in the ``loop unswitching'' optimization pass~\cite{llvmpasses}.
However, \papertitle{}'s invariant branch hoisting goes further by integrating loop unswitching with partial dead code elimination (PDCE)~\cite{knoop1994partial} to create additional opportunities for hoisting invariant branches further up.  
In this context, data cached by the DMA operations, such as \texttt{A} and \texttt{B} in Fig.~\ref{fig:tiropt}(d), become partially dead outside the \texttt{boundary(i)} condition, since the TIR lowering pass enforces that all the consumers of the loop are under the loop's constraint, i.e., the boundary check~\cite{feng2023tensorir}. 
Once PDCE moves these dependent DMA instructions under the boundary check (step \textcircled{2}), \papertitle{} can hoist the invariant branch even higher, outside the loop \texttt{j} (step \textcircled{3}). 
Considering that PDCE is typically excluded from low-level compilers due to its time complexity~\cite{knoop1994partial}, \papertitle{}'s invariant branch hoisting demonstrates the effectiveness of applying aggressive optimizations at a high level tailored to specialized algorithms and hardware. 
While this optimization does not eliminate or modify the boundary check, it significantly reduces the number of its dynamic instances (by 40$\times$) and those of DMA and compute operations (by 2$\times$).

\subsection{UPMEM Backend and Runtime Support}
\label{sec:drampim_runtime}

We extend the TVM C/C++ backend to support UPMEM as target hardware for final code generation from TIR programs to target-specific executables. 
The backend generates kernel C code with UPMEM built-in functions from optimized TIR programs and handles host code generation by lowering data transfer intrinsics in Section~\ref{subsec:tir_lowering} depending on access patterns. 
For constant tensors (e.g., weight matrix) reused throughout kernel execution, we allow users to guide the runtime to perform data transfers once before kernel launches, while data transfer intrinsics for vector data are directly compiled to loops with runtime function calls in the host code. 
Then, we reuse the TIR to LLVM IR lowering pass as it does not require UPMEM-specific handling and can be compiled to any hardware supported by the LLVM compiler~\cite{lattner2004llvm}.
\papertitle{} also provides runtime support for UPMEM host and kernel codes through the TVM runtime. These runtime interfaces include host APIs for UPMEM memory allocation/deallocation, data transfers and synchronizations between the host and DPUs, and resource acquisition for compute units, which are implemented using device driver API provided by the UPMEM SDK.
\section{Methodology}
\label{sec:methodology}

We implemented \papertitle{} based on the Apache TVM compiler framework~\cite{chen2018tvm} version 0.13.0 (commit 97c5de6). The TVM runtime for UPMEM was developed using ``Host and DPU Runtime Library'' functions provided by the UPMEM SDK version 2021.3.0.
For UPMEM kernel compilation, we used the \texttt{dpu-upmem-dpurte-clang} compiler from the SDK, which extends Clang-LLVM compiler version 12~\cite{upmemllvm}. Kernels were compiled with the \texttt{-O2} optimization flag.

\noindent
{\bf UPMEM server, simulator, and benchmark configurations.}
We used a UPMEM server with dual-socket Intel Xeon Gold 5220R CPUs and 32 ranks of PIM-enabled DIMMs totaling 2048 DPUs (banks) for code development and experiments. 
For an instruction-level performance breakdown for DPU, which is not available through standard UPMEM SDK profiling tools, we used uPIMulator~\cite{upimulator} (commit: 870d916).

We evaluated seven tensor algebra operations and 
two types of neural network layers, fully-connected (FC) layers and multi-head attention (MHA) layers, that represent the main target operations for PIM for their high memory intensity.
Tensor algebra operations are listed below:

\aptLtoX[graphic=no,type=html]{\begin{itemize}
    \item Vector addition (VA): {$C(i) = A(i) + B(i)$}
    \item Reduction (RED): {$b = \sum_{i} A(i)$}
    \item Matrix times vector (MTV): {$C(i) = A(i, j) \cdot B(j)$}

    \item Tensor times vector (TTV): {$C(i, j) = A(i, j, k) \cdot B(k)$}
    \item {Multiple matrix times vector (MMTV): $C(i, j) = \sum_{k} A(i, j, k) \cdot B(i, k)$}
    \item General vector addition (GEVA): {$C(i) = c \cdot A(i) + d \cdot B(i)$}
    \item {General matrix-vector multiplication (GEMV): $C(i) = c \cdot A(i, j) \cdot B(j)$}
\end{itemize}}{\begin{itemize}[leftmargin=*,parsep=0in,topsep=0.05in]
    \item Vector addition (VA): \textls[-10]{$C(i) = A(i) + B(i)$}
    \item Reduction (RED): \textls[-10]{$b = \sum_{i} A(i)$}
    \item Matrix times vector (MTV): \textls[-10]{$C(i) = A(i, j) \cdot B(j)$}

    \item Tensor times vector (TTV): \textls[-10]{$C(i, j) = A(i, j, k) \cdot B(k)$}
    \item \textls[-12]{Multiple matrix times vector (MMTV): $C(i, j) = \sum_{k} A(i, j, k) \cdot B(i, k)$}
    \item General vector addition (GEVA): \textls[-10]{$C(i) = c \cdot A(i) + d \cdot B(i)$}
    \item \textls[-12]{General matrix-vector multiplication (GEMV): $C(i) = c \cdot A(i, j) \cdot B(j)$}
\end{itemize}}

We also evaluated the fully connected (MTV) and MMTV operations of the multi-head attention layers of GPT-J 6B and 30B~\cite{ben2021gptj}, where these operations dominate runtime.
The FC layer includes four types of MTV operations: QKV generation, QKV projection, FC, and FC projection. MMTV operations are in the shape of (\# batch $\times$ \# heads, \# tokens, 256), with GPT-J 6B using 16 heads and GPT-J 13B using 28 heads. We performed experiments with batch sizes of 1, 4, and 16, and token sizes of 64, 128, 256, and 512 to cover common request sizes used in LLM datasets~\cite{dev2023sharegpt,taori2023alpaca}.

\noindent
{\bf Experimental setup.}
We used the PrIM benchmark~\cite{juan2022prim} as our baseline, designed for UPMEM hardware benchmarking and featuring a wide range of highly hand-optimized kernels.
We used VA, RED, and MTV from PrIM~\cite{juan2022prim} and wrote GEVA, TTV, MMTV, and GEMV based on PrIM's codes and programming methodology (PrIM-style). We modified PrIM's VA and MTV to add constant factors for GEVA and GEMV. For TTV, we flattened the outer loop dimension on the host side while using the unmodified MTV kernel for DPU. For MMTV, we adjusted the host code to distribute DPUs across the outer loop dimension.

The following configurations were used to evaluate and compare \papertitle{}'s efficiency with prior work in Section~\ref{subsection:upmem_with_autotuning}. 

\begin{itemize}[leftmargin=*,parsep=0in,topsep=0.05in]
    \item \textit{PrIM/PrIM(E):} 
    PrIM and PrIM-style benchmarks using default parameters from PrIM~\cite{juan2022prim}, except for the number of DPUs, which was selected via grid search ($2^n$ where $5 \leq n \leq 11$ for MMTV and $8 \leq n \leq 11$ for the rest).
    \item \textit{PrIM+search:} 
    PrIM and PrIM-style benchmarks, with the number of DPUs, the number of tasklets, and the WRAM caching tile size determined via grid search. This configuration highlights the importance of parameter searches in general and contrasts independent search spaces with \papertitle{}'s joint search space.
    \item \textit{SimplePIM:} VA and RED kernels provided by SimplePIM~\cite{jinfan2023simplepim}. 
    \item \textit{\papertitle{}:} Codes compiled by \papertitle{}'s search-based code generator as described in Section~\ref{sec:search_based_code_generation}. 
    \item \textit{CPU-autotuned:}
    Tensor programs autotuned for CPU using TVM's MetaSchedule~\cite{shao2022metaschedule} with default configurations.
\end{itemize}

\section{Experimental Results}
\label{sec:evaluation}

We evaluated the performance of \papertitle{}-compiled tensor operations on UPMEM and compared it with related work, focusing on the effectiveness of search-based code generation with the joint search space, the impact of each autotuning parameter on workload performance, and the advantages of tensor-level boundary check optimizations.
In summary, our experimental results demonstrate: 
\begin{itemize}[leftmargin=*,parsep=0in,topsep=0.05in]
\item Tensor operations auto-tuned with \papertitle{} outperform PrIM, PrIM(E), and SimplePIM by 2.49$\times$, 1.85$\times$, and 2.86$\times$ on average, respectively. Even when compared to PrIM+search, \papertitle{} achieves an average speedup of 1.84$\times$ by exploring a broader search space.
\item{\papertitle{}'s tensor-level boundary check optimizations effectively handle diverse kernels with varying tensor shapes, delivering performance comparable to or up to 20.5\% better than hand-tuned codes (PrIM). }
\item ATiM outperforms auto-tuned CPU versions for larger tensor sizes up to 23.3$\times$, for which its performance gains increase significantly due to reduced data movement overhead.
\end{itemize}

\subsection{Autotuned Performance of Tensor Programs}
\label{subsection:upmem_with_autotuning}

\begin{figure*}[!t]
    \centering
    \includegraphics[width=\textwidth]{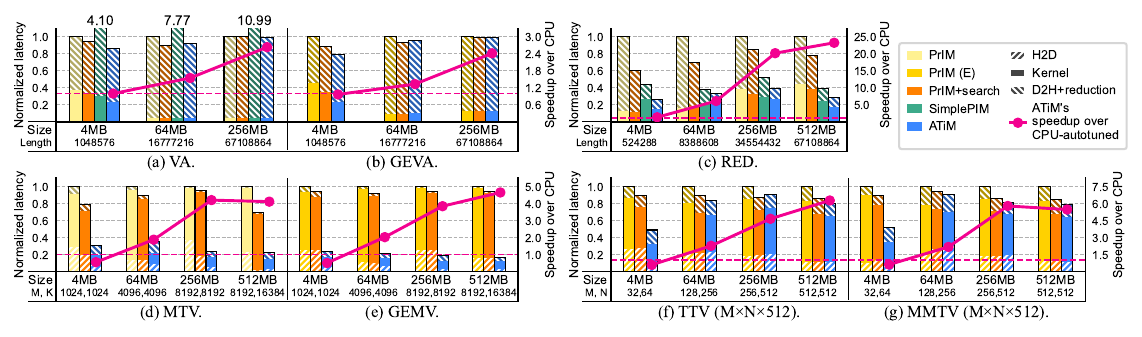}
    \vspace{-2.3\baselineskip}
    \caption{Tensor operation performance (normalized to PrIM).}
    \vspace{-0.7\baselineskip}
    \label{fig:polybench_results}
\end{figure*}

\begin{scriptsize}
\begin{table}[!t]
\centering
\caption{Autotuned parameters for PrIM, PrIM(E), PrIM+search and \papertitle{}. \papertitle{} specifies the number of DPUs per dimension type (spatial, reduction), and PrIM distributes DPUs only on the outermost spatial dimension.}
\vspace{-1.0\baselineskip}
\label{table:tuned_parameters}
\renewcommand{\arraystretch}{1.1}
\begin{tabular}{|m{0.095\linewidth}|m{0.05\linewidth}|m{0.05\linewidth}|m{0.05\linewidth}|m{0.06\linewidth}|m{0.085\linewidth}|m{0.07\linewidth}|m{0.06\linewidth}|m{0.09\linewidth}|}
\hline
{}&{} & \textbf{PrIM} & \multicolumn{3}{c|}{\textbf{PrIM+search}} & \multicolumn{3}{c|}{\textbf{ATiM}} \\
\cline{3-9}
\textbf{Workload} & \textbf{Size (MB)} & \textbf{DPUs} & \textbf{DPUs} &\textbf{\hspace*{-0.5\tabcolsep}tasklets} & \textbf{Caching tile size} & \textbf{DPUs /dim} &\textbf{\hspace*{-0.5\tabcolsep}tasklets} & \textbf{Caching tile size} \\
\hline
\multirow{4}{*}{\textbf{RED}}   & 4 & 256  & 256  & 8  & 32  & 128  & 8  & 32  \\
               & 64 & 1024 & 1024 & 8  & 256 & 512  & 16 & 64  \\
               & 256 & 1024 & 2048 & 16 & 256 & 1024 & 16 & 128 \\
               & 512 & 1024 & 2048 & 8  & 128 & 2048 & 16 & 128 \\
\hline
\multirow{4}{*}{\textbf{MTV}}   & 4 & 256  & 256  & 24 & 16  & (16,16)   & 16 & (32,32,4) \\
               & 64 & 256  & 256  & 16 & 16  & (32,64)   & 16 & (64,8,8)  \\
               & 256 & 512  & 256  & 24 & 8   & (128,16)  & 16 & (8,8,2)   \\
               & 512 & 512  & 512  & 24 & 16  & (32,64)   & 16 & (64,32,2) \\
\hline
\multirow{4}{*}{\textbf{GEMV}}  & 4 & 256  & 256  & 8  & 16  & (16,16)   & 16 & (64,32,2) \\
               & 64 & 256  & 256  & 8  & 16  & (256,8)   & 8  & (128,64,2) \\
               & 256 & 512  & 512  & 8  & 16  & (32,64)   & 16 & (128,32,8) \\
               & 512 & 512  & 512  & 8  & 16  & (128,16)  & 16 & (32,16,4) \\
\hline
\multirow{4}{*}{\textbf{TTV}}   & 4 & 256  & 256  & 8  & 16  & (32,8)    & 16 & (16,16,4) \\
               & 64 & 1024 & 1024 & 16 & 16  & (128,8)   & 16 & (64,32,2) \\
               & 256 & 2048 & 2048 & 16 & 16  & (2048,1)  & 16 & (64,64,4) \\
               & 512 & 2048 & 2048 & 24 & 8   & (512,4)   & 16 & (128,16,8) \\
\hline
\multirow{4}{*}{\textbf{MMTV}}  & 4 & 64   & 64   & 16 & 16  & (32,8)    & 16 & (128,32,4) \\
               & 64 & 512  & 512  & 24 & 16  & (2048,1)  & 16 & (128,64,2) \\
               & 256 & 2048 & 2048 & 16 & 16  & (2048,1)  & 16 & (16,16,2)  \\
               & 512 & 2048 & 2048 & 16 & 128 & (2048,1)  & 16 & (128,128,4)\\
\hline
\multirow{3}{*}{\textbf{VA}}    & 4 & 2048 & 2048 & 8  & 16  & 1024     & 8  & (32,32)   \\
               & 64 & 2048 & 2048 & 16 & 32  & 2048     & 8  & (256,256) \\
               & 256 & 2048 & 2048 & 16 & 64  & 2048     & 16 & (32,32)   \\
\hline
\multirow{3}{*}{\textbf{GEVA}}  & 4 & 1024 & 1024 & 8  & 128 & 1024     & 16 & (64,64)   \\
               & 64 & 1024 & 2048 & 24 & 32  & 2048     & 8  & (32,32)   \\
               & 256 & 2048 & 2048 & 16 & 16  & 2048     & 16 & (16,16)   \\
\hline
\end{tabular}
\vspace{-1.9\baselineskip}
\end{table}
\end{scriptsize}

\noindent
{\bf VA and GEVA. } 
\papertitle{} improves the performance of VA and GEVA by reducing kernel execution time by 28.2\% and 17.8\% on average, respectively. While PrIM follows the programming guide's recommendation of using 1024 bytes caching tiles, PrIM+search and \papertitle{} typically select smaller tiles between 64 and 256 bytes identified from searches.
Using smaller caching tiles prevents tasklets from becoming idle, thereby enhancing tasklet-level parallelism and more effectively hiding WRAM access latency through pipelined tasklet execution.
As shown in Fig.~\ref{fig:polybench_results}(a) and (b), the performance impact is more visible for smaller tensors, as D2H latencies dominate the runtime for larger tensors.
\papertitle{} further improves kernel performance by 13.2\% compared to PrIM+search by generating intra-DPU DMA instructions with static sizes, thus reducing the number of instructions needed to initiate DMA transfers. 
In contrast, SimplePIM performs 4-11$\times$ worse than PrIM and \papertitle{} due to inefficient D2H transfers, unnecessarily copying the entire tensor on the host side. 

\noindent
{\bf RED. } 
For the reduction operation, \papertitle{} significantly outperforms PrIM, PrIM+search, and SimplePIM by 3.19$\times$, 2.31$\times$, and 1.37$\times$ on average, respectively, by generating more efficient D2H data transfer and parallel reduction codes. 
When only one partial reduction result per DPU needs to be transferred to the host, PrIM and PrIM+search redundantly send the results from all tasklets, leading to excessive D2H data movement overhead.
While SimplePIM avoids redundant transfers, we found that its DPU partial reduction and host final reduction have inefficient implementations; the host final reduction code incurs extra overhead from invoking multiple internal library functions, while global barriers at each partial reduction step create higher synchronization overhead than the two-thread handshake mechanism used by PrIM(+search) and \papertitle{}.

\noindent
{\bf MTV and GEMV. } 
\papertitle{} achieves speedups up to 6.18$\times$ and 5.79$\times$ for MTV and GEMV over PrIM and PrIM+search. 
These performance gains result from a synergetic interplay of multiple autotuning factors, primarily 2D tiling with hierarchical reduction and caching tile location and size optimizations.
MTV and GEMV operations have a spatial loop dimension of up to 8192 elements, and if the tiling is performed only on this spatial dimension, there is insufficient inter-DPU parallelism to fully utilize 2048 DPUs. By applying 2D tiling on both spatial and reduction loop dimensions, \papertitle{} generates a sufficient number of smaller tiles to be distributed and computed, significantly reducing the kernel execution time up to 8.84$\times$. As shown in Table~\ref{table:tuned_parameters}, \papertitle{} identifies optimal 2D tile sizes (factors between 16 and 256) through autotuning, whereas the prior work relies only on 1D tiling. 
This also reduces H2D data transfer overheads, with \papertitle{} providing up to 8.37$\times$ speedup compared to baseline while slightly increasing post-processing time due to hierarchical reduction. 
Although it is challenging to isolate individual factors since multiple autotuning parameters interact with each other, the performance gap between PrIM and PrIM+search, both without reduction loop tiling, indicates that caching tile sizes also have a significant performance impact.

\begin{figure*}[!th]
    \centering
    \includegraphics[width=\textwidth]{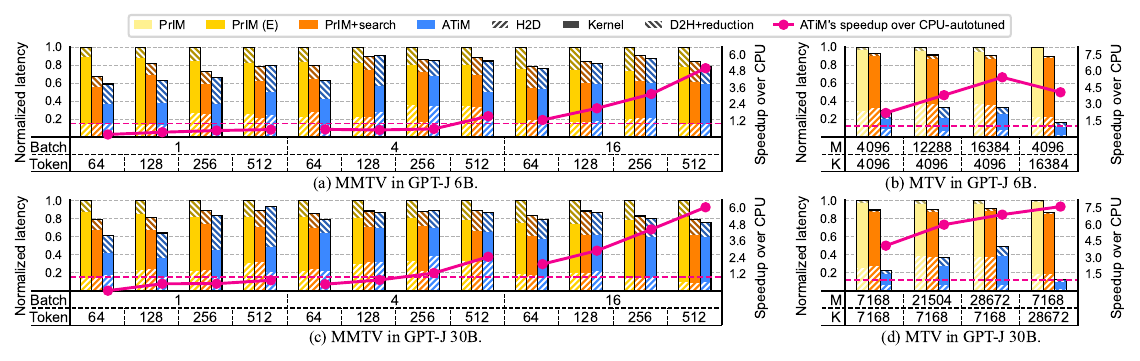}
    \vspace{-2.5\baselineskip}
    \caption{Performance of FC (MTV) and MMTV operations of the MHA layers of GPT-J 6B and 30B (normalized to PrIM). }
    \vspace{-1.2\baselineskip}
    \label{fig:gpt_results}
\end{figure*}

\noindent
{\bf TTV and MMTV. } 
TTV and MMTV achieve 2.04$\times$ and 1.94$\times$ speedups for small tensors, similar to MTV and GEMV, though the performance gains noticeably diminish with large tensors. 
While these kernels have more spatial loop levels than MTV and GEMV, providing greater inter-DPU and intra-DPU parallelism, the potential for tiling on the reduction loop dimension is limited by smaller reduction loop iterations. As a result, \papertitle{} makes similar DPU distribution decisions as PrIM(+search), exploiting spatial loop tiling only without parallel reduction. \papertitle{}'s balanced evolutionary search mechanism ensures schedule candidates without \texttt{rfactor} are not dropped early in the autotuning process in this case. 
On the other hand, for smaller tensors, where the reduction dimension relatively larger than the spatial dimensions, the benefit of reducing data movement through parallel reduction outweighs its overhead. Thus, \papertitle{} leverages additional parallelism on the reduction loop dimension to further improve H2D and kernel execution time by 1.44$\times$ and 3.02$\times$, respectively. 

\noindent
{\bf UPMEM vs. CPU.} 
Comparing five PIM versions, including ATiM, with an autotuned CPU version for each tensor program, we found that PIM consistently outperform CPU by up to 23.3$\times$ for tensors $\geq$ \SI{64}{MB}, 
while for smaller \SI{4}{MB} tensors, most PIM versions perform worse than CPUs due to high parallel overheads, except for RED, VA, and GEVA where ATiM performs comparably.
For all tensor programs, CPU versions suffer from increased data movement overhead as tensor size grows, while PIM versions benefit from in-place parallel computation across hundreds to thousands of DPUs. 
For MTV and GEMV, the data movement savings for PrIM(+search) are offset by limited inter-DPU parallelism. Its 1D tiling strategy, which tiles tensor rows only, fails to fully utilize parallel DPUs and generates relatively long kernels with large tiles. In contrast, ATiM significantly outperforms both CPU-autotuned and PrIM(+search) versions by identifying optimal 2D tiling configurations, fully leveraging DPU parallelism and hierarchical reduction.  

\begin{figure}[t]
    \centering
    \includegraphics[width=\linewidth]{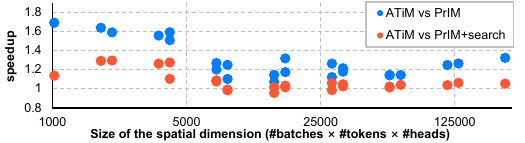}
    \vspace{-1.6\baselineskip}
    \caption{\papertitle{}'s speedup for MMTV with varying spatial dimension sizes. } 
    \label{fig:gpt_mmtv} 
    \vspace{-1.6\baselineskip}
\end{figure}

\subsection{Autotuned Performance of GPT-J Layers}
\label{subsection:upmem_with_autotuning_gpt}
We evaluated \papertitle{}'s performance on the MTV and MMTV from a real-world model (GPT-J) to assess its practical applicability (Fig.~\ref{fig:gpt_results}).

\noindent
{\bf MMTV in GPT-J.}
GPT-J has a fixed reduction loop size of 256 elements, but its spatial dimension as a product of the number of heads (depending on the model type), batch size, and number of tokens can vary significantly. 
Fig.~\ref{fig:gpt_results} illustrates \papertitle{}'s autotuning efficiency across MMTV kernels with different spatial dimension sizes, showing speedups ranging from 7.24\% to 69.1\%. 
GPT-6B, with fewer heads, outperforms GPT-30B by 5.72\% on average, and both models perform better with a single batch size compared to batch sizes of four and sixteen, achieving average speedups of 20.8\% and 11.7\%, respectively. 
As discussed earlier, with smaller spatial dimensions, the inter- and intra-DPU parallelism benefit from reduction dimension tiling becomes more pronounced.  
Fig.~\ref{fig:gpt_mmtv} plotting the relationship between spatial dimension sizes and speedups shows that \papertitle{} offers significant advantages over PrIM(+search) for spatial dimensions $\leq$ 5000, after which speedups plateau. 
Additionally, \papertitle{}'s fine-grained caching size autotuning ability provides further speedups beyond reduction dimension tiling.
For example, with GPT-J 6B (batch size of 4 and 64 tokens), \papertitle{} achieves a 59.2\% speedup by using two distinct caching tile sizes found through searches, while PrIM+search uses one size.

\noindent
{\bf MTV in GPT-J.} 
\papertitle{} shows significant performance improvements over PrIM and PrIM+search (up to 8.21$\times$ and 7.11$\times$) across all four MTV kernels in GPT-J. As discussed in the performance analysis for the MTV kernel, tiling on the reduction loop dimension (column) effectively reduces both kernel execution and data transfer times. 
Moreover, the benefits of reduction loop tiling increase as the reduction dimension becomes larger relative to the spatial dimension (row), offsetting post-reduction overhead. For instance, \papertitle{} achieves a 3.03$\times$ speedup for a 16384$\times$4096 tensor, compared to a 6.25$\times$ speedup for its transposed form, 4096$\times$16384.

\begin{figure*}[ht]
    \centering
    \includegraphics[width=\linewidth]{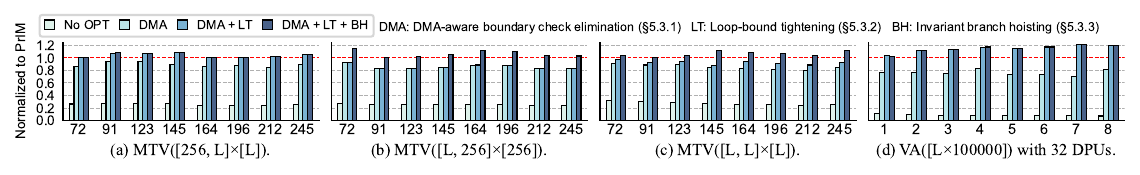}
    \vspace{-2.2\baselineskip}
    \caption{Kernel performance with PIM-aware optimizations in Section~\ref{sec:kernel_code_optimization} applied. (a) and (b) are misaligned on one axis, (c) on both axes, and (d) misaligned VA.   }
    \label{fig:opt_experiment} 
    \vspace{-1.2\baselineskip}
\end{figure*}

\noindent
{\bf UPMEM vs. CPU.} 
For the MMTV operations in the MHA layer, ATiM achieves speedups of up to 5.01$\times$ and 6.07$\times$ over CPU for the 6B and 30B parameter models, respectively.
Notably, while ATiM-autotuned versions initially perform slower than CPU-autotuned versions with smaller workloads (up to a batch size of 4 and 256 tokens), they show significant performance improvements beyond this point, scaling proportionally with increasing tensor sizes.  
We observed that for small tensors, communication reduction overheads outweigh parallelization benefits as the number of DPUs increases. Using fewer DPUs can reduce these overheads but limits performance due to reduced inter-DPU parallelism. Conversely, with larger tensor dimensions, UPMEM scales more effectively than the CPU by fully exploiting data parallelism across DPUs and better amortizing parallel overheads.

The MTV workload consistently outperforms the CPU across all tensor shapes, achieving speedups of up to 7.65$\times$. Since each tensor size is at least 64MB in this scenario, DRAM bandwidth becomes the primary bottleneck for the CPU. ATiM not only mitigates this bottleneck through in-memory processing but also leverages higher inter-DPU parallelism, delivering greater and more scalable acceleration. Both ATiM and CPU show performance variations for the tensors of the same size with different (transposed) shapes, as shown in the last two columns in Fig.~\ref{fig:gpt_results}(b) and (d). 

\subsection{PIM-aware Optimization Performance}
\label{subsection:upmem_effect_pim_aware_optim}

Fig.~\ref{fig:opt_experiment} shows the performance of the MTV and VA kernels optimized by different combinations of PIM-aware optimizations in Section~\ref{sec:kernel_code_optimization}, compared to PrIM versions. When all three optimizations are applied, MTV and VA achieve up to 14.7\% and 20.5\% (5.37\% and 14.4\% on average), respectively. 
For all scenarios, DMA-aware boundary check elimination (\texttt{DMA}) provides the largest performance gain by replacing element-wise memory assignment and boundary checks with MRAM-WRAM DMA instructions. Loop-bound tightening (\texttt{LT}) further improves performance for workloads with misaligned columns (Fig.~\ref{fig:opt_experiment}(a)) by eliminating redundant boundary checks inside computation loops (\texttt{k} loop in Fig.~\ref{fig:tiropt}). 
Invariant branch hoisting (\texttt{BH}) optimizes tensors with misaligned rows (Fig.~\ref{fig:opt_experiment}(b)) by lifting invariant branches with row variables (\texttt{i} in Fig.~\ref{fig:tiropt}) outside main loops and skipping unnecessary iterations. 
For workloads with both row and column misalignments, such as dynamic square shapes in Fig.~\ref{fig:opt_experiment}(c), both \texttt{LT} and \texttt{BH} are applied to reduce branch overheads on all axes. 
These optimizations are also effective for other kernels; for example, \texttt{LT} reduces redundant branches in the VA kernel's computation loop, achieving up to a 20.5\% speedup (Fig.~\ref{fig:opt_experiment}(d)). In summary, \papertitle{} effectively optimizes loops with boundary checks, generating kernel codes with minimal branch penalties for UPMEM.

\begin{figure}[t]
    \centering
    \includegraphics[width=\linewidth]{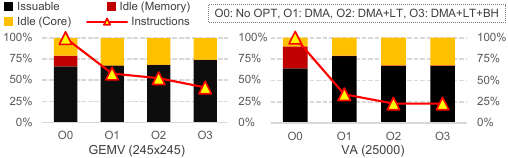}
    \vspace{-2.0\baselineskip}
    \caption{Breakdown of a single \mbox{DPU’s} runtime (active in black and idle in red and yellow) under PIM-aware optimizations, as simulated with uPIMulator. The line with triangular markers indicates the number of instructions, normalized to \mbox{\texttt{No-OPT}} (\texttt{O0}).}
    \label{fig:opt_analysis} 
    \vspace{-1.2\baselineskip}
\end{figure}

\begin{figure}[t]
    \centering
    \includegraphics[width=\linewidth]{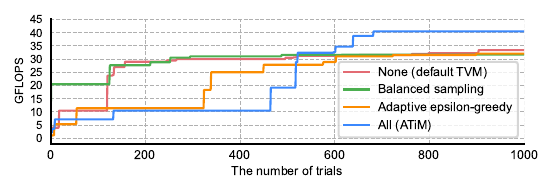}
    \vspace{-2.5\baselineskip}
    \caption{Autotuning efficiency of balanced sampling and adaptive epsilon-greedy strategies (individually and in combination).}
    \label{fig:gflops_plot} 
    \vspace{-1.2\baselineskip}
\end{figure}

We performed additional simulator-based experiments to further analyze how these optimizations impact data movement and kernel execution performance (Fig.~\mbox{\ref{fig:opt_analysis}}). The results confirm our assumption that frequent, small DMA requests in \mbox{\texttt{O0}} incur significant memory stalls, which are eliminated by \mbox{DMA‐aware boundary‐check} elimination (\mbox{\texttt{O1}}). The subsequent \mbox{loop‐bound} tightening (\mbox{\texttt{O2}}) and invariant branch hoisting (\mbox{\texttt{O3}}) optimizations further decrease the total instruction count
by reducing the number of DMA requests, branches, and loop iterations.

\subsection{Balanced Evolutionary Search Result}
\label{subsection:balanced_evolutionary_search}

The combination of balanced sampling and adaptive epsilon-greedy strategy, as detailed in Section~\ref{subsection:balanced_evolutionary_search}, improves the autotuning process to converge at a higher performance result than TVM's original evolutionary search or standalone application of either technique.
While \papertitle{} initially shows lower autotuning efficiency, its balanced exploration of \texttt{rfactor} and \texttt{non-rfactor} candidates allows it to converge to optimal performance after 400 trials (40\% of the total), delivering significant speedups—21.2\% over TVM, and 26.4\% and 27.9\% over each standalone version (Fig.~\ref{fig:gflops_plot}).
When applied individually, each optimization may penalize early iterations without addressing candidate imbalances, ultimately performing similarly to the default TVM.

\section{Discussion}
\label{sec:discussion}

\begin{figure}[t]
    \centering
    \includegraphics[width=\linewidth]{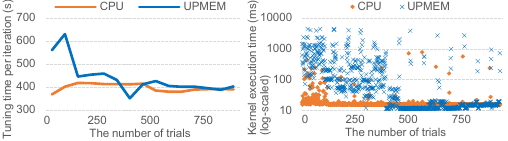}
    \vspace{-1.8\baselineskip}
    \caption{Per-iteration autotuning time (left) and candidate execution time (right) throughout 16 iterations (64 trials each) on CPU and UPMEM (ATiM).}
    \label{fig:autotuning_overhead} 
    \vspace{-1.2\baselineskip}
\end{figure}

\noindent 
{\bf Autotuning overheads. }
Compared to the CPU autotuning time with the original TVM compiler, the UPMEM autotuning time with ATiM is 11.5\% and 18.5\% higher for the MTV and MMTV kernels with 4, 64, 256, and \SI{512}{MB} tensors for 1000 trials. 
A detailed analysis of the autotuning time throughout iterations reveals that ATiM spends significantly more time on hardware measurements than the CPU while it eventually converges to a better-performing version. ATiM's balanced evolutionary search prioritizes exploration in the early autotuning iterations to cover a larger search space than CPUs, which results in higher initial autotuning overheads (left graph of Fig.~\ref{fig:autotuning_overhead}). Additionally, we observed that UPMEM shows greater performance variability with suboptimal DPU tiling configurations, while CPU performance remains relatively stable, as shown in the outlier candidates (blue dots in the right graph of Fig.~\ref{fig:autotuning_overhead}). We plan to address these issues by refining the search algorithm and cost model in future work.  

\noindent
{\bf Extension to other DRAM-PIM architectures. }
ATiM is designed to support DRAM-PIM architectures with bank-level compute units by exploiting tensor-level IR for common abstractions and specializing the TIR lowering passes for specific hardware targets. For DRAM-PIM with MAC acceleration logic instead of general-purpose processors~\cite{lee2021hbmpim,lee2022gddr6aim}, ATiM's autotuner and code generator would need to be extended to handle different bank mapping policies and compile to PIM-specific vector intrinsics or tensorized functions.  
For instance, HBM-PIM assigns a processing unit (PU) to every two banks and activates it with memory commands in a special mode to perform matrix-vector operations with data from memory rows and PU registers~\cite{lee2021hbmpim}. To support this architecture, ATiM must redefine the DPU binding in two levels -- one at the bank level and another at the PU level -- and support specialized vector intrinsic using TVM's tensorization feature~\cite{chen2018tvm} to enforce compatible loop structures. Our preliminary results using an open-source HBM-PIM simulator~\cite{dev2022pimsimulator} and library~\cite{dev2023pimlibrary} suggest that ATiM can be seamlessly extended to support an automated host and kernel code generation for HBM-PIM. 
Future work includes evolving ATiM into an extensible and portable tensor compiler for diverse  DRAM-PIM hardware.

\noindent 
{\bf DL framework interfaces. }
To interface with DL frameworks and automate end-to-end compilation from input models, it is necessary to extend the graph-level TVM frontend using Relax~\cite{lai2023relax} or Relay~\cite{roesch2018relay} IR to invoke ATiM and update the TVM runtime to manage layer and memory mapping to DRAM-PIM hardware.
Once trained models are imported from other DL frameworks, such as TensorFlow~\cite{abadi2016tensorflow}, PyTorch~\cite{paszke2019pytorch}, and ONNX~\cite{bai2019onnx}, by the TVM frontend, ATiM can further perform inter-kernel optimizations at the graph IR level, e.g., layer fusion and reordering, to reduce inter-DPU data transfers and DPU kernel launch overheads.
Integrating ATiM with application interfaces and optimizing its system-level usage is an important direction for our future work.

\noindent
{\bf Integration with MLIR.}
MLIR~\cite{lattner2021mlir} can serve as an alternative high-level IR for PIM compilers~\cite{khan2023cinm}. While ATiM relies on TVM's autotuning framework, its core techniques, such as schedule primitives for data distribution and kernel optimizations, can be incorporated into MLIR's transform dialect~\cite{lucke2024mlirtransformdialect}, which enables dialect-to-dialect transformations. For example, GPU programs in MLIR GPU dialects (e.g., Triton~\cite{triton}) could be lowered and optimized for PIM execution by defining transformation rules. Fully integrating ATiM with MLIR would require independent research, involving the development of cost-based search algorithms and significant dialect extensions.

\noindent
{\bf Broader application of DMA and boundary optimizations. }
The PIM-aware optimizations in Section \ref{sec:kernel_code_optimization} can be generalized beyond UPMEM, benefiting hardware with similar architectural characteristics.
Specifically, DMA-aware boundary check elimination, which enables safe data overfetching via DMA, aligns well with accelerators that support efficient bulk memory transfers~\cite{vogel2018fpga, codrescu2014hexagon, versalacap}.
Loop-bound tightening and invariant branch hoisting (Sections \ref{subsection:loopextent} and \ref{subsection:hoist}) can effectively improve performance for resource-constrained processors, such as RISC-V and ARM-based edge CPUs, which are widely used in embedded AI and IoT devices yet lack advanced branch prediction and other latency-hiding mechanisms~\cite{sifive, asanovic2016rocket, yiu2009cortexm3}.
As future work, we plan to develop abstractions and algorithms to enable more portable application of these optimizations across a broader range of hardware targets.

\section{Related Work}
\label{sec:related_work}

\noindent
{\bf Tensor program support for DRAM-PIM.}
Recent research has actively explored software support for DRAM-PIM hardware~\cite{jinfan2023simplepim,lin2023songc,shin2023pimflow,oliveria2024mimdram,khan2023cinm,ibrahim2024balanced,gu2020ipim,kwon2022aimsoftware,sait2022onemcc}. Vendor-provided solutions~\cite{sait2022onemcc, kwon2022aimsoftware} offer full software stacks but rely heavily on hand-optimized libraries, highlighting the need for optimizing code generators like \papertitle{} for tensor programs.
Several efforts proposed DRAM-PIM specific compilation and optimization support but have limitations compared to \papertitle{}.
For instance, CINM~\cite{khan2023cinm} defines MLIR dialects for various PIM architectures but lacks systematic optimization and auto-tuning capabilities. iPIM~\cite{gu2020ipim} employs a scheduling language for efficient data mapping but supports limited data distribution strategies tailored to the cube-structured memory of HMC~\cite{pawlowski2011hmc}. SimplePIM~\cite{jinfan2023simplepim} provides primitives for UPMEM but is restricted to 1D arrays, limiting its support for general tensor operations. PIMFlow~\cite{shin2023pimflow} extends TVM for CNN models on DRAM-PIM but is not fully integrated with TVM's runtime and autotuner, relying on fixed optimization strategies.
Other studies, TC-CIM~\cite{drebes2020tccim}, target specific PIM hardware like Computing-in-Memory (CIM), focusing on tensor mapping that considers its analog computation nature. In contrast, ATiM jointly optimizes host and kernel operations for DRAM-PIM architectures, offering a more flexible and comprehensive approach. 

\noindent
{\bf PIM-specific code optimizations.}
Prior work has shown that host data distribution and kernel multi-level tiling are crucial for achieving high performance in DRAM-PIM systems~\cite{giannoula2022sparsep, ibrahim2024balanced, dietcode}. 
Studies such as \cite{khan2023cinm, giannoula2022sparsep} revealed that efficient data distribution can reduce data movement and enhance DPU parallelism, improving UPMEM performance. While these works explored a limited set of kernel optimization schemes, \papertitle{} systematically explores the host and kernel loop optimization space.
PIMnast~\cite{ibrahim2024balanced} examined performance-tuning factors for DRAM-PIM hardware, highlighting how memory configuration and operation requirements influence data distribution strategies and kernel optimizations through tiling and caching strategies. Building upon these insights, \papertitle{} implements a search-based code generation mechanism with PIM-specific autotuning parameters.
Additionally, DietCode~\cite{dietcode} proposed algorithm-level optimizations to reduce boundary check overheads, some of which have been incorporated into TVM.
\papertitle{} leverages these techniques to introduce more aggressive optimizations tailored to UPMEM hardware constraints and tensor loop structures.
\section{Conclusion and Future Work}
\label{sec:conclusion}

\papertitle{} tackles the challenge of building a fully automated code generation support for emerging DRAM-centric accelerators while efficiently navigating the vast space of host and kernel optimizations. 
It carefully expands existing tensor compiler infrastructure to provide flexible, autotuning-driven code generation for UPMEM, complying with its tensor language semantics and seamlessly integrating with its lowering pipeline. Experimental results show \papertitle{}'s strong potential as a practical, optimizing tensor compiler for diverse DRAM-PIM architectures. For our future work, we will focus on developing \papertitle{}'s application-level interfaces, supporting other types of DRAM-PIM hardware, and enabling graph-level autotuning.

\begin{acks}
We thank the anonymous reviewers for their valuable feedback. 
This work was supported in part by Samsung Advanced Institute of Technology (SAIT) and in part by Institute for Information \& communications Technology Planning \& Evaluation (IITP) grants (RS-2019-II191906
, 2021-0-00871
, 2021-0-00310
, RS-2023-00228970
) and National Research Foundation (NRF) grants (%
RS-2023-00222663
, RS-2023-00283799
) funded by the Korean government (MSIT).
\end{acks}

\bibliographystyle{ACM-Reference-Format}
\bibliography{references}


\begin{thebibliography}{67}


\ifx \showCODEN    \undefined \def \showCODEN     #1{\unskip}     \fi
\ifx \showDOI      \undefined \def \showDOI       #1{#1}\fi
\ifx \showISBNx    \undefined \def \showISBNx     #1{\unskip}     \fi
\ifx \showISBNxiii \undefined \def \showISBNxiii  #1{\unskip}     \fi
\ifx \showISSN     \undefined \def \showISSN      #1{\unskip}     \fi
\ifx \showLCCN     \undefined \def \showLCCN      #1{\unskip}     \fi
\ifx \shownote     \undefined \def \shownote      #1{#1}          \fi
\ifx \showarticletitle \undefined \def \showarticletitle #1{#1}   \fi
\ifx \showURL      \undefined \def \showURL       {\relax}        \fi
\providecommand\bibfield[2]{#2}
\providecommand\bibinfo[2]{#2}
\providecommand\natexlab[1]{#1}
\providecommand\showeprint[2][]{arXiv:#2}

\bibitem[Abadi et~al\mbox{.}(2016)]%
        {abadi2016tensorflow}
\bibfield{author}{\bibinfo{person}{Mart{\'\i}n Abadi}, \bibinfo{person}{Paul Barham}, \bibinfo{person}{Jianmin Chen}, \bibinfo{person}{Zhifeng Chen}, \bibinfo{person}{Andy Davis}, \bibinfo{person}{Jeffrey Dean}, \bibinfo{person}{Matthieu Devin}, \bibinfo{person}{Sanjay Ghemawat}, \bibinfo{person}{Geoffrey Irving}, \bibinfo{person}{Michael Isard}, {et~al\mbox{.}}} \bibinfo{year}{2016}\natexlab{}.
\newblock \showarticletitle{$\{$TensorFlow$\}$: a system for $\{$Large-Scale$\}$ machine learning}. In \bibinfo{booktitle}{\emph{12th USENIX symposium on operating systems design and implementation (OSDI 16)}}. \bibinfo{pages}{265--283}.
\newblock


\bibitem[Asanovic et~al\mbox{.}(2016)]%
        {asanovic2016rocket}
\bibfield{author}{\bibinfo{person}{Krste Asanovic}, \bibinfo{person}{Rimas Avizienis}, \bibinfo{person}{Jonathan Bachrach}, \bibinfo{person}{Scott Beamer}, \bibinfo{person}{David Biancolin}, \bibinfo{person}{Christopher Celio}, \bibinfo{person}{Henry Cook}, \bibinfo{person}{Daniel Dabbelt}, \bibinfo{person}{John Hauser}, \bibinfo{person}{Adam Izraelevitz}, {et~al\mbox{.}}} \bibinfo{year}{2016}\natexlab{}.
\newblock \showarticletitle{The rocket chip generator}.
\newblock \bibinfo{journal}{\emph{EECS Department, University of California, Berkeley, Tech. Rep. UCB/EECS-2016-17}}  \bibinfo{volume}{4} (\bibinfo{year}{2016}), \bibinfo{pages}{6--2}.
\newblock


\bibitem[Bai et~al\mbox{.}(2019)]%
        {bai2019onnx}
\bibfield{author}{\bibinfo{person}{Junjie Bai}, \bibinfo{person}{Fang Lu}, \bibinfo{person}{Ke Zhang}, {et~al\mbox{.}}} \bibinfo{year}{2019}\natexlab{}.
\newblock \bibinfo{title}{{ONNX: Open Neural Network Exchange}}.
\newblock \bibinfo{howpublished}{\url{https://github.com/onnx/onnx}}.
\newblock


\bibitem[Bernhardt et~al\mbox{.}(2023)]%
        {bernhardt2023pimdb}
\bibfield{author}{\bibinfo{person}{Arthur Bernhardt}, \bibinfo{person}{Andreas Koch}, {and} \bibinfo{person}{Ilia Petrov}.} \bibinfo{year}{2023}\natexlab{}.
\newblock \showarticletitle{pimDB: From Main-Memory DBMS to Processing-In-Memory DBMS-Engines on Intelligent Memories}. In \bibinfo{booktitle}{\emph{Proceedings of the 19th International Workshop on Data Management on New Hardware}} (Seattle, WA, USA) \emph{(\bibinfo{series}{DaMoN '23})}. \bibinfo{publisher}{Association for Computing Machinery}, \bibinfo{address}{New York, NY, USA}, \bibinfo{pages}{44–52}.
\newblock
\showISBNx{9798400701917}
\urldef\tempurl%
\url{https://doi.org/10.1145/3592980.3595312}
\showDOI{\tempurl}


\bibitem[Chen et~al\mbox{.}(2023a)]%
        {jinfan2023simplepim}
\bibfield{author}{\bibinfo{person}{J. Chen}, \bibinfo{person}{J. Gomez-Luna}, \bibinfo{person}{I.~El Hajj}, \bibinfo{person}{Y. Guo}, {and} \bibinfo{person}{O. Mutlu}.} \bibinfo{year}{2023}\natexlab{a}.
\newblock \showarticletitle{SimplePIM: A Software Framework for Productive and Efficient Processing-in-Memory}. In \bibinfo{booktitle}{\emph{2023 32nd International Conference on Parallel Architectures and Compilation Techniques (PACT)}}. \bibinfo{publisher}{IEEE Computer Society}, \bibinfo{address}{Los Alamitos, CA, USA}, \bibinfo{pages}{99--111}.
\newblock
\urldef\tempurl%
\url{https://doi.org/10.1109/PACT58117.2023.00017}
\showDOI{\tempurl}


\bibitem[Chen et~al\mbox{.}(2023b)]%
        {chen2023uppipe}
\bibfield{author}{\bibinfo{person}{Liang-Chi Chen}, \bibinfo{person}{Chien-Chung Ho}, {and} \bibinfo{person}{Yuan-Hao Chang}.} \bibinfo{year}{2023}\natexlab{b}.
\newblock \showarticletitle{UpPipe: A Novel Pipeline Management on In-Memory Processors for RNA-seq Quantification}. In \bibinfo{booktitle}{\emph{2023 60th ACM/IEEE Design Automation Conference (DAC)}}. \bibinfo{pages}{1--6}.
\newblock
\urldef\tempurl%
\url{https://doi.org/10.1109/DAC56929.2023.10247915}
\showDOI{\tempurl}


\bibitem[Chen et~al\mbox{.}(2018a)]%
        {chen2018tvm}
\bibfield{author}{\bibinfo{person}{Tianqi Chen}, \bibinfo{person}{Thierry Moreau}, \bibinfo{person}{Ziheng Jiang}, \bibinfo{person}{Lianmin Zheng}, \bibinfo{person}{Eddie Yan}, \bibinfo{person}{Haichen Shen}, \bibinfo{person}{Meghan Cowan}, \bibinfo{person}{Leyuan Wang}, \bibinfo{person}{Yuwei Hu}, \bibinfo{person}{Luis Ceze}, \bibinfo{person}{Carlos Guestrin}, {and} \bibinfo{person}{Arvind Krishnamurthy}.} \bibinfo{year}{2018}\natexlab{a}.
\newblock \showarticletitle{{TVM}: An Automated {End-to-End} Optimizing Compiler for Deep Learning}. In \bibinfo{booktitle}{\emph{13th USENIX Symposium on Operating Systems Design and Implementation (OSDI 18)}}. \bibinfo{publisher}{USENIX Association}, \bibinfo{address}{Carlsbad, CA}, \bibinfo{pages}{578--594}.
\newblock
\showISBNx{978-1-939133-08-3}
\urldef\tempurl%
\url{https://www.usenix.org/conference/osdi18/presentation/chen}
\showURL{%
\tempurl}


\bibitem[Chen et~al\mbox{.}(2018b)]%
        {chen1028autotvm}
\bibfield{author}{\bibinfo{person}{Tianqi Chen}, \bibinfo{person}{Lianmin Zheng}, \bibinfo{person}{Eddie Yan}, \bibinfo{person}{Ziheng Jiang}, \bibinfo{person}{Thierry Moreau}, \bibinfo{person}{Luis Ceze}, \bibinfo{person}{Carlos Guestrin}, {and} \bibinfo{person}{Arvind Krishnamurthy}.} \bibinfo{year}{2018}\natexlab{b}.
\newblock \showarticletitle{Learning to Optimize Tensor Programs}. In \bibinfo{booktitle}{\emph{Advances in Neural Information Processing Systems}}, \bibfield{editor}{\bibinfo{person}{S.~Bengio}, \bibinfo{person}{H.~Wallach}, \bibinfo{person}{H.~Larochelle}, \bibinfo{person}{K.~Grauman}, \bibinfo{person}{N.~Cesa-Bianchi}, {and} \bibinfo{person}{R.~Garnett}} (Eds.), Vol.~\bibinfo{volume}{31}. \bibinfo{publisher}{Curran Associates, Inc.}
\newblock
\urldef\tempurl%
\url{https://proceedings.neurips.cc/paper_files/paper/2018/file/8b5700012be65c9da25f49408d959ca0-Paper.pdf}
\showURL{%
\tempurl}


\bibitem[Cho et~al\mbox{.}(2021)]%
        {cho2021accel}
\bibfield{author}{\bibinfo{person}{Benjamin~Y. Cho}, \bibinfo{person}{Jeageun Jung}, {and} \bibinfo{person}{Mattan Erez}.} \bibinfo{year}{2021}\natexlab{}.
\newblock \showarticletitle{Accelerating bandwidth-bound deep learning inference with main-memory accelerators}. In \bibinfo{booktitle}{\emph{Proceedings of the International Conference for High Performance Computing, Networking, Storage and Analysis}} (St. Louis, Missouri) \emph{(\bibinfo{series}{SC '21})}. \bibinfo{publisher}{Association for Computing Machinery}, \bibinfo{address}{New York, NY, USA}, Article \bibinfo{articleno}{44}, \bibinfo{numpages}{14}~pages.
\newblock
\showISBNx{9781450384421}
\urldef\tempurl%
\url{https://doi.org/10.1145/3458817.3476146}
\showDOI{\tempurl}


\bibitem[Codrescu et~al\mbox{.}(2014)]%
        {codrescu2014hexagon}
\bibfield{author}{\bibinfo{person}{Lucian Codrescu}, \bibinfo{person}{Willie Anderson}, \bibinfo{person}{Suresh Venkumanhanti}, \bibinfo{person}{Mao Zeng}, \bibinfo{person}{Erich Plondke}, \bibinfo{person}{Chris Koob}, \bibinfo{person}{Ajay Ingle}, \bibinfo{person}{Charles Tabony}, {and} \bibinfo{person}{Rick Maule}.} \bibinfo{year}{2014}\natexlab{}.
\newblock \showarticletitle{Hexagon DSP: An architecture optimized for mobile multimedia and communications}.
\newblock \bibinfo{journal}{\emph{IEEE Micro}} \bibinfo{volume}{34}, \bibinfo{number}{2} (\bibinfo{year}{2014}), \bibinfo{pages}{34--43}.
\newblock


\bibitem[Das et~al\mbox{.}(2022)]%
        {das2022dl}
\bibfield{author}{\bibinfo{person}{Prangon Das}, \bibinfo{person}{Purab~Ranjan Sutradhar}, \bibinfo{person}{Mark Indovina}, \bibinfo{person}{Sai Manoj~Pudukotai Dinakarrao}, {and} \bibinfo{person}{Amlan Ganguly}.} \bibinfo{year}{2022}\natexlab{}.
\newblock \showarticletitle{Implementation and Evaluation of Deep Neural Networks in Commercially Available Processing in Memory Hardware}. In \bibinfo{booktitle}{\emph{2022 IEEE 35th International System-on-Chip Conference (SOCC)}}. \bibinfo{pages}{1--6}.
\newblock
\urldef\tempurl%
\url{https://doi.org/10.1109/SOCC56010.2022.9908126}
\showDOI{\tempurl}


\bibitem[Devaux(2019a)]%
        {devaux2019upmem}
\bibfield{author}{\bibinfo{person}{Fabrice Devaux}.} \bibinfo{year}{2019}\natexlab{a}.
\newblock \showarticletitle{The true Processing In Memory accelerator}. In \bibinfo{booktitle}{\emph{2019 IEEE Hot Chips 31 Symposium (HCS)}}. \bibinfo{pages}{1--24}.
\newblock
\urldef\tempurl%
\url{https://doi.org/10.1109/HOTCHIPS.2019.8875680}
\showDOI{\tempurl}


\bibitem[Devaux(2019b)]%
        {fabrice2019upmem}
\bibfield{author}{\bibinfo{person}{Fabrice Devaux}.} \bibinfo{year}{2019}\natexlab{b}.
\newblock \showarticletitle{The true Processing In Memory accelerator}. In \bibinfo{booktitle}{\emph{2019 IEEE Hot Chips 31 Symposium (HCS)}}. \bibinfo{pages}{1--24}.
\newblock
\urldef\tempurl%
\url{https://doi.org/10.1109/HOTCHIPS.2019.8875680}
\showDOI{\tempurl}


\bibitem[Draper et~al\mbox{.}(2002)]%
        {jeff2002diva}
\bibfield{author}{\bibinfo{person}{Jeff Draper}, \bibinfo{person}{Jacqueline Chame}, \bibinfo{person}{Mary Hall}, \bibinfo{person}{Craig Steele}, \bibinfo{person}{Tim Barrett}, \bibinfo{person}{Jeff LaCoss}, \bibinfo{person}{John Granacki}, \bibinfo{person}{Jaewook Shin}, \bibinfo{person}{Chun Chen}, \bibinfo{person}{Chang~Woo Kang}, \bibinfo{person}{Ihn Kim}, {and} \bibinfo{person}{Gokhan Daglikoca}.} \bibinfo{year}{2002}\natexlab{}.
\newblock \showarticletitle{The architecture of the DIVA processing-in-memory chip}. In \bibinfo{booktitle}{\emph{Proceedings of the 16th International Conference on Supercomputing}} (New York, New York, USA) \emph{(\bibinfo{series}{ICS '02})}. \bibinfo{publisher}{Association for Computing Machinery}, \bibinfo{address}{New York, NY, USA}, \bibinfo{pages}{14–25}.
\newblock
\showISBNx{1581134835}
\urldef\tempurl%
\url{https://doi.org/10.1145/514191.514197}
\showDOI{\tempurl}


\bibitem[Drebes et~al\mbox{.}(2020)]%
        {drebes2020tccim}
\bibfield{author}{\bibinfo{person}{Andi Drebes}, \bibinfo{person}{Lorenzo Chelini}, \bibinfo{person}{Oleksandr Zinenko}, \bibinfo{person}{Albert Cohen}, \bibinfo{person}{Henk Corporaal}, \bibinfo{person}{Tobias Grosser}, \bibinfo{person}{Kanishkan Vadivel}, {and} \bibinfo{person}{Nicolas Vasilache}.} \bibinfo{year}{2020}\natexlab{}.
\newblock \showarticletitle{TC-CIM: Empowering Tensor Comprehensions for Computing-In-Memory}.
\newblock
\urldef\tempurl%
\url{http://impact.gforge.inria.fr/impact2020/}
\showURL{%
\tempurl}
\newblock
\shownote{10th International Workshop on Polyhedral Compilation Techniques, IMPACT 2010 ; Conference date: 22-01-2020 Through 22-01-2020}.


\bibitem[Feng et~al\mbox{.}(2023)]%
        {feng2023tensorir}
\bibfield{author}{\bibinfo{person}{Siyuan Feng}, \bibinfo{person}{Bohan Hou}, \bibinfo{person}{Hongyi Jin}, \bibinfo{person}{Wuwei Lin}, \bibinfo{person}{Junru Shao}, \bibinfo{person}{Ruihang Lai}, \bibinfo{person}{Zihao Ye}, \bibinfo{person}{Lianmin Zheng}, \bibinfo{person}{Cody~Hao Yu}, \bibinfo{person}{Yong Yu}, {and} \bibinfo{person}{Tianqi Chen}.} \bibinfo{year}{2023}\natexlab{}.
\newblock \showarticletitle{TensorIR: An Abstraction for Automatic Tensorized Program Optimization}. In \bibinfo{booktitle}{\emph{Proceedings of the 28th ACM International Conference on Architectural Support for Programming Languages and Operating Systems, Volume 2}} (Vancouver, BC, Canada) \emph{(\bibinfo{series}{ASPLOS 2023})}. \bibinfo{publisher}{Association for Computing Machinery}, \bibinfo{address}{New York, NY, USA}, \bibinfo{pages}{804–817}.
\newblock
\showISBNx{9781450399166}
\urldef\tempurl%
\url{https://doi.org/10.1145/3575693.3576933}
\showDOI{\tempurl}


\bibitem[Giannoula et~al\mbox{.}(2022)]%
        {giannoula2022sparsep}
\bibfield{author}{\bibinfo{person}{Christina Giannoula}, \bibinfo{person}{Ivan Fernandez}, \bibinfo{person}{Juan Gómez-Luna}, \bibinfo{person}{Nectarios Koziris}, \bibinfo{person}{Georgios Goumas}, {and} \bibinfo{person}{Onur Mutlu}.} \bibinfo{year}{2022}\natexlab{}.
\newblock \showarticletitle{SparseP: Efficient Sparse Matrix Vector Multiplication on Real Processing-In-Memory Architectures}. In \bibinfo{booktitle}{\emph{2022 IEEE Computer Society Annual Symposium on VLSI (ISVLSI)}}. \bibinfo{pages}{288--291}.
\newblock
\urldef\tempurl%
\url{https://doi.org/10.1109/ISVLSI54635.2022.00063}
\showDOI{\tempurl}


\bibitem[Gogineni et~al\mbox{.}(2024)]%
        {gogineni2024swiftrl}
\bibfield{author}{\bibinfo{person}{Kailash Gogineni}, \bibinfo{person}{Sai~Santosh Dayapule}, \bibinfo{person}{Juan Gómez-Luna}, \bibinfo{person}{Karthikeya Gogineni}, \bibinfo{person}{Peng Wei}, \bibinfo{person}{Tian Lan}, \bibinfo{person}{Mohammad Sadrosadati}, \bibinfo{person}{Onur Mutlu}, {and} \bibinfo{person}{Guru Venkataramani}.} \bibinfo{year}{2024}\natexlab{}.
\newblock \showarticletitle{SwiftRL: Towards Efficient Reinforcement Learning on Real Processing-In-Memory Systems}. In \bibinfo{booktitle}{\emph{2024 IEEE International Symposium on Performance Analysis of Systems and Software (ISPASS)}}. \bibinfo{pages}{217--229}.
\newblock
\urldef\tempurl%
\url{https://doi.org/10.1109/ISPASS61541.2024.00029}
\showDOI{\tempurl}


\bibitem[Gokhale et~al\mbox{.}(1995)]%
        {gokhale1995pim}
\bibfield{author}{\bibinfo{person}{M. Gokhale}, \bibinfo{person}{B. Holmes}, {and} \bibinfo{person}{K. Iobst}.} \bibinfo{year}{1995}\natexlab{}.
\newblock \showarticletitle{Processing in memory: the Terasys massively parallel PIM array}.
\newblock \bibinfo{journal}{\emph{Computer}} \bibinfo{volume}{28}, \bibinfo{number}{4} (\bibinfo{year}{1995}), \bibinfo{pages}{23--31}.
\newblock
\urldef\tempurl%
\url{https://doi.org/10.1109/2.375174}
\showDOI{\tempurl}


\bibitem[Grosser et~al\mbox{.}(2012)]%
        {grosser2012polly}
\bibfield{author}{\bibinfo{person}{Tobias Grosser}, \bibinfo{person}{Armin Groesslinger}, {and} \bibinfo{person}{Christian Lengauer}.} \bibinfo{year}{2012}\natexlab{}.
\newblock \showarticletitle{Polly—performing polyhedral optimizations on a low-level intermediate representation}.
\newblock \bibinfo{journal}{\emph{Parallel Processing Letters}} \bibinfo{volume}{22}, \bibinfo{number}{04} (\bibinfo{year}{2012}), \bibinfo{pages}{1250010}.
\newblock


\bibitem[Gu et~al\mbox{.}(2020)]%
        {gu2020ipim}
\bibfield{author}{\bibinfo{person}{Peng Gu}, \bibinfo{person}{Xinfeng Xie}, \bibinfo{person}{Yufei Ding}, \bibinfo{person}{Guoyang Chen}, \bibinfo{person}{Weifeng Zhang}, \bibinfo{person}{Dimin Niu}, {and} \bibinfo{person}{Yuan Xie}.} \bibinfo{year}{2020}\natexlab{}.
\newblock \showarticletitle{iPIM: Programmable In-Memory Image Processing Accelerator Using Near-Bank Architecture}. In \bibinfo{booktitle}{\emph{2020 ACM/IEEE 47th Annual International Symposium on Computer Architecture (ISCA)}}. \bibinfo{pages}{804--817}.
\newblock
\urldef\tempurl%
\url{https://doi.org/10.1109/ISCA45697.2020.00071}
\showDOI{\tempurl}


\bibitem[Gómez-Luna et~al\mbox{.}(2023)]%
        {2023gomezlunaupmemtraining}
\bibfield{author}{\bibinfo{person}{Juan Gómez-Luna}, \bibinfo{person}{Yuxin Guo}, \bibinfo{person}{Sylvan Brocard}, \bibinfo{person}{Julien Legriel}, \bibinfo{person}{Remy Cimadomo}, \bibinfo{person}{Geraldo~F. Oliveira}, \bibinfo{person}{Gagandeep Singh}, {and} \bibinfo{person}{Onur Mutlu}.} \bibinfo{year}{2023}\natexlab{}.
\newblock \showarticletitle{Evaluating Machine LearningWorkloads on Memory-Centric Computing Systems}. In \bibinfo{booktitle}{\emph{2023 IEEE International Symposium on Performance Analysis of Systems and Software (ISPASS)}}. \bibinfo{pages}{35--49}.
\newblock
\urldef\tempurl%
\url{https://doi.org/10.1109/ISPASS57527.2023.00013}
\showDOI{\tempurl}


\bibitem[Gómez-Luna et~al\mbox{.}(2022)]%
        {juan2022prim}
\bibfield{author}{\bibinfo{person}{Juan Gómez-Luna}, \bibinfo{person}{Izzat~El Hajj}, \bibinfo{person}{Ivan Fernandez}, \bibinfo{person}{Christina Giannoula}, \bibinfo{person}{Geraldo~F. Oliveira}, {and} \bibinfo{person}{Onur Mutlu}.} \bibinfo{year}{2022}\natexlab{}.
\newblock \showarticletitle{Benchmarking a New Paradigm: Experimental Analysis and Characterization of a Real Processing-in-Memory System}.
\newblock \bibinfo{journal}{\emph{IEEE Access}}  \bibinfo{volume}{10} (\bibinfo{year}{2022}), \bibinfo{pages}{52565--52608}.
\newblock
\urldef\tempurl%
\url{https://doi.org/10.1109/ACCESS.2022.3174101}
\showDOI{\tempurl}


\bibitem[Hyun et~al\mbox{.}(2024)]%
        {upimulator}
\bibfield{author}{\bibinfo{person}{Bongjoon Hyun}, \bibinfo{person}{Taehun Kim}, \bibinfo{person}{Dongjae Lee}, {and} \bibinfo{person}{Minsoo Rhu}.} \bibinfo{year}{2024}\natexlab{}.
\newblock \showarticletitle{Pathfinding Future PIM Architectures by Demystifying a Commercial PIM Technology}. In \bibinfo{booktitle}{\emph{2024 IEEE International Symposium on High-Performance Computer Architecture (HPCA)}}. IEEE, \bibinfo{pages}{263--279}.
\newblock


\bibitem[Ibrahim et~al\mbox{.}(2024)]%
        {ibrahim2024balanced}
\bibfield{author}{\bibinfo{person}{Mohamed~Assem Ibrahim}, \bibinfo{person}{Mahzabeen Islam}, {and} \bibinfo{person}{Shaizeen Aga}.} \bibinfo{year}{2024}\natexlab{}.
\newblock \bibinfo{title}{Balanced Data Placement for GEMV Acceleration with Processing-In-Memory}.
\newblock
\newblock
\showeprint[arxiv]{2403.20297}~[cs.AR]


\bibitem[Khan et~al\mbox{.}(2023)]%
        {khan2023cinm}
\bibfield{author}{\bibinfo{person}{Asif~Ali Khan}, \bibinfo{person}{Hamid Farzaneh}, \bibinfo{person}{Karl F.~A. Friebel}, \bibinfo{person}{Clément Fournier}, \bibinfo{person}{Lorenzo Chelini}, {and} \bibinfo{person}{Jeronimo Castrillon}.} \bibinfo{year}{2023}\natexlab{}.
\newblock \bibinfo{title}{CINM (Cinnamon): A Compilation Infrastructure for Heterogeneous Compute In-Memory and Compute Near-Memory Paradigms}.
\newblock
\newblock
\showeprint[arxiv]{2301.07486}~[cs.AR]


\bibitem[Kim et~al\mbox{.}(2022)]%
        {kim2022lpddr5pim}
\bibfield{author}{\bibinfo{person}{Jin~Hyun Kim}, \bibinfo{person}{Shin-Haeng Kang}, \bibinfo{person}{Sukhan Lee}, \bibinfo{person}{Hyeonsu Kim}, \bibinfo{person}{Yuhwan Ro}, \bibinfo{person}{Seungwon Lee}, \bibinfo{person}{David Wang}, \bibinfo{person}{Jihyun Choi}, \bibinfo{person}{Jinin So}, \bibinfo{person}{YeonGon Cho}, \bibinfo{person}{JoonHo Song}, \bibinfo{person}{Jeonghyeon Cho}, \bibinfo{person}{Kyomin Sohn}, {and} \bibinfo{person}{Nam~Sung Kim}.} \bibinfo{year}{2022}\natexlab{}.
\newblock \showarticletitle{Aquabolt-XL HBM2-PIM, LPDDR5-PIM With In-Memory Processing, and AXDIMM With Acceleration Buffer}.
\newblock \bibinfo{journal}{\emph{IEEE Micro}} \bibinfo{volume}{42}, \bibinfo{number}{3} (\bibinfo{year}{2022}), \bibinfo{pages}{20--30}.
\newblock
\urldef\tempurl%
\url{https://doi.org/10.1109/MM.2022.3164651}
\showDOI{\tempurl}


\bibitem[Kim et~al\mbox{.}(2021)]%
        {kim2021aquabolt}
\bibfield{author}{\bibinfo{person}{Jin~Hyun Kim}, \bibinfo{person}{Shin-haeng Kang}, \bibinfo{person}{Sukhan Lee}, \bibinfo{person}{Hyeonsu Kim}, \bibinfo{person}{Woongjae Song}, \bibinfo{person}{Yuhwan Ro}, \bibinfo{person}{Seungwon Lee}, \bibinfo{person}{David Wang}, \bibinfo{person}{Hyunsung Shin}, \bibinfo{person}{Bengseng Phuah}, \bibinfo{person}{Jihyun Choi}, \bibinfo{person}{Jinin So}, \bibinfo{person}{YeonGon Cho}, \bibinfo{person}{JoonHo Song}, \bibinfo{person}{Jangseok Choi}, \bibinfo{person}{Jeonghyeon Cho}, \bibinfo{person}{Kyomin Sohn}, \bibinfo{person}{Youngsoo Sohn}, \bibinfo{person}{Kwangil Park}, {and} \bibinfo{person}{Nam~Sung Kim}.} \bibinfo{year}{2021}\natexlab{}.
\newblock \showarticletitle{Aquabolt-XL: Samsung HBM2-PIM with in-memory processing for ML accelerators and beyond}. In \bibinfo{booktitle}{\emph{2021 IEEE Hot Chips 33 Symposium (HCS)}}. \bibinfo{pages}{1--26}.
\newblock
\urldef\tempurl%
\url{https://doi.org/10.1109/HCS52781.2021.9567191}
\showDOI{\tempurl}


\bibitem[Knoop et~al\mbox{.}(1994)]%
        {knoop1994partial}
\bibfield{author}{\bibinfo{person}{Jens Knoop}, \bibinfo{person}{Oliver R{\"u}thing}, {and} \bibinfo{person}{Bernhard Steffen}.} \bibinfo{year}{1994}\natexlab{}.
\newblock \showarticletitle{Partial dead code elimination}.
\newblock \bibinfo{journal}{\emph{ACM Sigplan Notices}} \bibinfo{volume}{29}, \bibinfo{number}{6} (\bibinfo{year}{1994}), \bibinfo{pages}{147--158}.
\newblock


\bibitem[Kruse and Finkel(2018)]%
        {clangdirectives}
\bibfield{author}{\bibinfo{person}{Michael Kruse} {and} \bibinfo{person}{Hal Finkel}.} \bibinfo{year}{2018}\natexlab{}.
\newblock \showarticletitle{User-Directed Loop-Transformations in Clang}. In \bibinfo{booktitle}{\emph{2018 IEEE/ACM 5th Workshop on the LLVM Compiler Infrastructure in HPC (LLVM-HPC)}}. \bibinfo{pages}{49--58}.
\newblock
\urldef\tempurl%
\url{https://doi.org/10.1109/LLVM-HPC.2018.8639402}
\showDOI{\tempurl}


\bibitem[Kwon et~al\mbox{.}(2022)]%
        {kwon2022aimsoftware}
\bibfield{author}{\bibinfo{person}{Yongkee Kwon}, \bibinfo{person}{Kornijcuk Vladimir}, \bibinfo{person}{Nahsung Kim}, \bibinfo{person}{Woojae Shin}, \bibinfo{person}{Jongsoon Won}, \bibinfo{person}{Minkyu Lee}, \bibinfo{person}{Hyunha Joo}, \bibinfo{person}{Haerang Choi}, \bibinfo{person}{Guhyun Kim}, \bibinfo{person}{Byeongju An}, \bibinfo{person}{Jeongbin Kim}, \bibinfo{person}{Jaewook Lee}, \bibinfo{person}{Ilkon Kim}, \bibinfo{person}{Jaehan Park}, \bibinfo{person}{Chanwook Park}, \bibinfo{person}{Yosub Song}, \bibinfo{person}{Byeongsu Yang}, \bibinfo{person}{Hyungdeok Lee}, \bibinfo{person}{Seho Kim}, \bibinfo{person}{Daehan Kwon}, \bibinfo{person}{Seongju Lee}, \bibinfo{person}{Kyuyoung Kim}, \bibinfo{person}{Sanghoon Oh}, \bibinfo{person}{Joonhong Park}, \bibinfo{person}{Gimoon Hong}, \bibinfo{person}{Dongyoon Ka}, \bibinfo{person}{Kyudong Hwang}, \bibinfo{person}{Jeongje Park}, \bibinfo{person}{Kyeongpil Kang}, \bibinfo{person}{Jungyeon Kim}, \bibinfo{person}{Junyeol Jeon}, \bibinfo{person}{Myeongjun
  Lee}, \bibinfo{person}{Minyoung Shin}, \bibinfo{person}{Minhwan Shin}, \bibinfo{person}{Jaekyung Cha}, \bibinfo{person}{Changson Jung}, \bibinfo{person}{Kijoon Chang}, \bibinfo{person}{Chunseok Jeong}, \bibinfo{person}{Euicheol Lim}, \bibinfo{person}{Il Park}, \bibinfo{person}{Junhyun Chun}, {and} \bibinfo{person}{Sk Hynix}.} \bibinfo{year}{2022}\natexlab{}.
\newblock \showarticletitle{System Architecture and Software Stack for GDDR6-AiM}. In \bibinfo{booktitle}{\emph{2022 IEEE Hot Chips 34 Symposium (HCS)}}. \bibinfo{pages}{1--25}.
\newblock
\urldef\tempurl%
\url{https://doi.org/10.1109/HCS55958.2022.9895629}
\showDOI{\tempurl}


\bibitem[Lai et~al\mbox{.}(2023)]%
        {lai2023relax}
\bibfield{author}{\bibinfo{person}{Ruihang Lai}, \bibinfo{person}{Junru Shao}, \bibinfo{person}{Siyuan Feng}, \bibinfo{person}{Steven~S Lyubomirsky}, \bibinfo{person}{Bohan Hou}, \bibinfo{person}{Wuwei Lin}, \bibinfo{person}{Zihao Ye}, \bibinfo{person}{Hongyi Jin}, \bibinfo{person}{Yuchen Jin}, \bibinfo{person}{Jiawei Liu}, {et~al\mbox{.}}} \bibinfo{year}{2023}\natexlab{}.
\newblock \showarticletitle{Relax: Composable Abstractions for End-to-End Dynamic Machine Learning}.
\newblock \bibinfo{journal}{\emph{arXiv preprint arXiv:2311.02103}} (\bibinfo{year}{2023}).
\newblock


\bibitem[Lattner and Adve(2004)]%
        {lattner2004llvm}
\bibfield{author}{\bibinfo{person}{C. Lattner} {and} \bibinfo{person}{V. Adve}.} \bibinfo{year}{2004}\natexlab{}.
\newblock \showarticletitle{LLVM: a compilation framework for lifelong program analysis \& transformation}. In \bibinfo{booktitle}{\emph{International Symposium on Code Generation and Optimization, 2004. CGO 2004.}} \bibinfo{pages}{75--86}.
\newblock
\urldef\tempurl%
\url{https://doi.org/10.1109/CGO.2004.1281665}
\showDOI{\tempurl}


\bibitem[Lattner et~al\mbox{.}(2021)]%
        {lattner2021mlir}
\bibfield{author}{\bibinfo{person}{Chris Lattner}, \bibinfo{person}{Mehdi Amini}, \bibinfo{person}{Uday Bondhugula}, \bibinfo{person}{Albert Cohen}, \bibinfo{person}{Andy Davis}, \bibinfo{person}{Jacques Pienaar}, \bibinfo{person}{River Riddle}, \bibinfo{person}{Tatiana Shpeisman}, \bibinfo{person}{Nicolas Vasilache}, {and} \bibinfo{person}{Oleksandr Zinenko}.} \bibinfo{year}{2021}\natexlab{}.
\newblock \showarticletitle{MLIR: Scaling Compiler Infrastructure for Domain Specific Computation}. In \bibinfo{booktitle}{\emph{2021 IEEE/ACM International Symposium on Code Generation and Optimization (CGO)}}. \bibinfo{pages}{2--14}.
\newblock
\urldef\tempurl%
\url{https://doi.org/10.1109/CGO51591.2021.9370308}
\showDOI{\tempurl}


\bibitem[Lee et~al\mbox{.}(2021)]%
        {lee2021hbmpim}
\bibfield{author}{\bibinfo{person}{Sukhan Lee}, \bibinfo{person}{Shin-haeng Kang}, \bibinfo{person}{Jaehoon Lee}, \bibinfo{person}{Hyeonsu Kim}, \bibinfo{person}{Eojin Lee}, \bibinfo{person}{Seungwoo Seo}, \bibinfo{person}{Hosang Yoon}, \bibinfo{person}{Seungwon Lee}, \bibinfo{person}{Kyounghwan Lim}, \bibinfo{person}{Hyunsung Shin}, \bibinfo{person}{Jinhyun Kim}, \bibinfo{person}{O Seongil}, \bibinfo{person}{Anand Iyer}, \bibinfo{person}{David Wang}, \bibinfo{person}{Kyomin Sohn}, {and} \bibinfo{person}{Nam~Sung Kim}.} \bibinfo{year}{2021}\natexlab{}.
\newblock \showarticletitle{Hardware Architecture and Software Stack for PIM Based on Commercial DRAM Technology : Industrial Product}. In \bibinfo{booktitle}{\emph{2021 ACM/IEEE 48th Annual International Symposium on Computer Architecture (ISCA)}}. \bibinfo{pages}{43--56}.
\newblock
\urldef\tempurl%
\url{https://doi.org/10.1109/ISCA52012.2021.00013}
\showDOI{\tempurl}


\bibitem[Lee et~al\mbox{.}(2022)]%
        {lee2022gddr6aim}
\bibfield{author}{\bibinfo{person}{Seongju Lee}, \bibinfo{person}{Kyuyoung Kim}, \bibinfo{person}{Sanghoon Oh}, \bibinfo{person}{Joonhong Park}, \bibinfo{person}{Gimoon Hong}, \bibinfo{person}{Dongyoon Ka}, \bibinfo{person}{Kyudong Hwang}, \bibinfo{person}{Jeongje Park}, \bibinfo{person}{Kyeongpil Kang}, \bibinfo{person}{Jungyeon Kim}, \bibinfo{person}{Junyeol Jeon}, \bibinfo{person}{Nahsung Kim}, \bibinfo{person}{Yongkee Kwon}, \bibinfo{person}{Kornijcuk Vladimir}, \bibinfo{person}{Woojae Shin}, \bibinfo{person}{Jongsoon Won}, \bibinfo{person}{Minkyu Lee}, \bibinfo{person}{Hyunha Joo}, \bibinfo{person}{Haerang Choi}, \bibinfo{person}{Jaewook Lee}, \bibinfo{person}{Donguc Ko}, \bibinfo{person}{Younggun Jun}, \bibinfo{person}{Keewon Cho}, \bibinfo{person}{Ilwoong Kim}, \bibinfo{person}{Choungki Song}, \bibinfo{person}{Chunseok Jeong}, \bibinfo{person}{Daehan Kwon}, \bibinfo{person}{Jieun Jang}, \bibinfo{person}{Il Park}, \bibinfo{person}{Junhyun Chun}, {and} \bibinfo{person}{Joohwan Cho}.}
  \bibinfo{year}{2022}\natexlab{}.
\newblock \showarticletitle{A 1ynm 1.25V 8Gb, 16Gb/s/pin GDDR6-based Accelerator-in-Memory supporting 1TFLOPS MAC Operation and Various Activation Functions for Deep-Learning Applications}. In \bibinfo{booktitle}{\emph{2022 IEEE International Solid-State Circuits Conference (ISSCC)}}, Vol.~\bibinfo{volume}{65}. \bibinfo{pages}{1--3}.
\newblock
\urldef\tempurl%
\url{https://doi.org/10.1109/ISSCC42614.2022.9731711}
\showDOI{\tempurl}


\bibitem[Lim et~al\mbox{.}(2023)]%
        {lim2023join}
\bibfield{author}{\bibinfo{person}{Chaemin Lim}, \bibinfo{person}{Suhyun Lee}, \bibinfo{person}{Jinwoo Choi}, \bibinfo{person}{Jounghoo Lee}, \bibinfo{person}{Seongyeon Park}, \bibinfo{person}{Hanjun Kim}, \bibinfo{person}{Jinho Lee}, {and} \bibinfo{person}{Youngsok Kim}.} \bibinfo{year}{2023}\natexlab{}.
\newblock \showarticletitle{Design and Analysis of a Processing-in-DIMM Join Algorithm: A Case Study with UPMEM DIMMs}.
\newblock \bibinfo{journal}{\emph{Proc. ACM Manag. Data}} \bibinfo{volume}{1}, \bibinfo{number}{2}, Article \bibinfo{articleno}{113} (\bibinfo{date}{June} \bibinfo{year}{2023}), \bibinfo{numpages}{27}~pages.
\newblock
\urldef\tempurl%
\url{https://doi.org/10.1145/3589258}
\showDOI{\tempurl}


\bibitem[Lin et~al\mbox{.}(2023)]%
        {lin2023songc}
\bibfield{author}{\bibinfo{person}{Junfeng Lin}, \bibinfo{person}{Huanyu Qu}, \bibinfo{person}{Songchen Ma}, \bibinfo{person}{Xinglong Ji}, \bibinfo{person}{Hongyi Li}, \bibinfo{person}{Xiaochuan Li}, \bibinfo{person}{Chenhang Song}, {and} \bibinfo{person}{Weihao Zhang}.} \bibinfo{year}{2023}\natexlab{}.
\newblock \showarticletitle{SongC: A Compiler for Hybrid Near-Memory and In-Memory Many-Core Architecture}.
\newblock \bibinfo{journal}{\emph{IEEE Trans. Comput.}} (\bibinfo{year}{2023}), \bibinfo{pages}{1--14}.
\newblock
\urldef\tempurl%
\url{https://doi.org/10.1109/TC.2023.3311948}
\showDOI{\tempurl}


\bibitem[LLVM({[n.\,d.]})]%
        {llvmpasses}
\bibfield{author}{\bibinfo{person}{LLVM}.} \bibinfo{year}{[n.\,d.]}\natexlab{}.
\newblock \bibinfo{title}{LLVM’s Analysis and Transform Passes}.
\newblock \bibinfo{howpublished}{\url{https://llvm.org/docs/Passes.html}}.
\newblock


\bibitem[L{\"u}cke et~al\mbox{.}(2024)]%
        {lucke2024mlirtransformdialect}
\bibfield{author}{\bibinfo{person}{Martin~Paul L{\"u}cke}, \bibinfo{person}{Oleksandr Zinenko}, \bibinfo{person}{William~S Moses}, \bibinfo{person}{Michel Steuwer}, {and} \bibinfo{person}{Albert Cohen}.} \bibinfo{year}{2024}\natexlab{}.
\newblock \showarticletitle{The MLIR Transform Dialect. Your compiler is more powerful than you think}.
\newblock \bibinfo{journal}{\emph{arXiv preprint arXiv:2409.03864}} (\bibinfo{year}{2024}).
\newblock


\bibitem[Mutlu et~al\mbox{.}(2022)]%
        {mutlu2022modern}
\bibfield{author}{\bibinfo{person}{Onur Mutlu}, \bibinfo{person}{Saugata Ghose}, \bibinfo{person}{Juan Gómez-Luna}, {and} \bibinfo{person}{Rachata Ausavarungnirun}.} \bibinfo{year}{2022}\natexlab{}.
\newblock \showarticletitle{A modern primer on processing in memory}.
\newblock In \bibinfo{booktitle}{\emph{Emerging Computing: From Devices to Systems: Looking Beyond Moore and Von Neumann}}. \bibinfo{publisher}{Springer}, \bibinfo{pages}{171--243}.
\newblock


\bibitem[Nair et~al\mbox{.}(2015)]%
        {nair2015memorycube}
\bibfield{author}{\bibinfo{person}{R. Nair}, \bibinfo{person}{S.~F. Antao}, \bibinfo{person}{C. Bertolli}, \bibinfo{person}{P. Bose}, \bibinfo{person}{J.~R. Brunheroto}, \bibinfo{person}{T. Chen}, \bibinfo{person}{C.-Y. Cher}, \bibinfo{person}{C.~H.~A. Costa}, \bibinfo{person}{J. Doi}, \bibinfo{person}{C. Evangelinos}, \bibinfo{person}{B.~M. Fleischer}, \bibinfo{person}{T.~W. Fox}, \bibinfo{person}{D.~S. Gallo}, \bibinfo{person}{L. Grinberg}, \bibinfo{person}{J.~A. Gunnels}, \bibinfo{person}{A.~C. Jacob}, \bibinfo{person}{P. Jacob}, \bibinfo{person}{H.~M. Jacobson}, \bibinfo{person}{T. Karkhanis}, \bibinfo{person}{C. Kim}, \bibinfo{person}{J.~H. Moreno}, \bibinfo{person}{J.~K. O'Brien}, \bibinfo{person}{M. Ohmacht}, \bibinfo{person}{Y. Park}, \bibinfo{person}{D.~A. Prener}, \bibinfo{person}{B.~S. Rosenburg}, \bibinfo{person}{K.~D. Ryu}, \bibinfo{person}{O. Sallenave}, \bibinfo{person}{M.~J. Serrano}, \bibinfo{person}{P.~D.~M. Siegl}, \bibinfo{person}{K. Sugavanam}, {and} \bibinfo{person}{Z. Sura}.}
  \bibinfo{year}{2015}\natexlab{}.
\newblock \showarticletitle{Active Memory Cube: A processing-in-memory architecture for exascale systems}.
\newblock \bibinfo{journal}{\emph{IBM Journal of Research and Development}} \bibinfo{volume}{59}, \bibinfo{number}{2/3} (\bibinfo{year}{2015}), \bibinfo{pages}{17:1--17:14}.
\newblock
\urldef\tempurl%
\url{https://doi.org/10.1147/JRD.2015.2409732}
\showDOI{\tempurl}


\bibitem[Nickolls et~al\mbox{.}(2008)]%
        {nickolls2008cuda}
\bibfield{author}{\bibinfo{person}{John Nickolls}, \bibinfo{person}{Ian Buck}, \bibinfo{person}{Michael Garland}, {and} \bibinfo{person}{Kevin Skadron}.} \bibinfo{year}{2008}\natexlab{}.
\newblock \showarticletitle{Scalable parallel programming with CUDA}. In \bibinfo{booktitle}{\emph{ACM SIGGRAPH 2008 Classes}} (Los Angeles, California) \emph{(\bibinfo{series}{SIGGRAPH '08})}. \bibinfo{publisher}{Association for Computing Machinery}, \bibinfo{address}{New York, NY, USA}, Article \bibinfo{articleno}{16}, \bibinfo{numpages}{14}~pages.
\newblock
\showISBNx{9781450378451}
\urldef\tempurl%
\url{https://doi.org/10.1145/1401132.1401152}
\showDOI{\tempurl}


\bibitem[Oliveira et~al\mbox{.}(2024)]%
        {oliveria2024mimdram}
\bibfield{author}{\bibinfo{person}{G.~F. Oliveira}, \bibinfo{person}{A. Olgun}, \bibinfo{person}{A. Yaglikci}, \bibinfo{person}{F. Bostanci}, \bibinfo{person}{J. Gomez-Luna}, \bibinfo{person}{S. Ghose}, {and} \bibinfo{person}{O. Mutlu}.} \bibinfo{year}{2024}\natexlab{}.
\newblock \showarticletitle{MIMDRAM: An End-to-End Processing-Using-DRAM System for High-Throughput, Energy-Efficient and Programmer-Transparent Multiple-Instruction Multiple-Data Computing}. In \bibinfo{booktitle}{\emph{2024 IEEE International Symposium on High-Performance Computer Architecture (HPCA)}}. \bibinfo{publisher}{IEEE Computer Society}, \bibinfo{address}{Los Alamitos, CA, USA}, \bibinfo{pages}{186--203}.
\newblock
\urldef\tempurl%
\url{https://doi.org/10.1109/HPCA57654.2024.00024}
\showDOI{\tempurl}


\bibitem[Paszke et~al\mbox{.}(2019)]%
        {paszke2019pytorch}
\bibfield{author}{\bibinfo{person}{Adam Paszke}, \bibinfo{person}{Sam Gross}, \bibinfo{person}{Francisco Massa}, \bibinfo{person}{Adam Lerer}, \bibinfo{person}{James Bradbury}, \bibinfo{person}{Gregory Chanan}, \bibinfo{person}{Trevor Killeen}, \bibinfo{person}{Zeming Lin}, \bibinfo{person}{Natalia Gimelshein}, \bibinfo{person}{Luca Antiga}, \bibinfo{person}{Alban Desmaison}, \bibinfo{person}{Andreas K\"{o}pf}, \bibinfo{person}{Edward Yang}, \bibinfo{person}{Zach DeVito}, \bibinfo{person}{Martin Raison}, \bibinfo{person}{Alykhan Tejani}, \bibinfo{person}{Sasank Chilamkurthy}, \bibinfo{person}{Benoit Steiner}, \bibinfo{person}{Lu Fang}, \bibinfo{person}{Junjie Bai}, {and} \bibinfo{person}{Soumith Chintala}.} \bibinfo{year}{2019}\natexlab{}.
\newblock \bibinfo{booktitle}{\emph{PyTorch: an imperative style, high-performance deep learning library}}.
\newblock \bibinfo{publisher}{Curran Associates Inc.}, \bibinfo{address}{Red Hook, NY, USA}.
\newblock


\bibitem[Pawlowski(2011)]%
        {pawlowski2011hmc}
\bibfield{author}{\bibinfo{person}{J.~Thomas Pawlowski}.} \bibinfo{year}{2011}\natexlab{}.
\newblock \showarticletitle{Hybrid memory cube (HMC)}. In \bibinfo{booktitle}{\emph{2011 IEEE Hot Chips 23 Symposium (HCS)}}. \bibinfo{pages}{1--24}.
\newblock
\urldef\tempurl%
\url{https://doi.org/10.1109/HOTCHIPS.2011.7477494}
\showDOI{\tempurl}


\bibitem[Ragan-Kelley et~al\mbox{.}(2013)]%
        {ragankelley2013halide}
\bibfield{author}{\bibinfo{person}{Jonathan Ragan-Kelley}, \bibinfo{person}{Connelly Barnes}, \bibinfo{person}{Andrew Adams}, \bibinfo{person}{Sylvain Paris}, \bibinfo{person}{Fr\'{e}do Durand}, {and} \bibinfo{person}{Saman Amarasinghe}.} \bibinfo{year}{2013}\natexlab{}.
\newblock \showarticletitle{Halide: a language and compiler for optimizing parallelism, locality, and recomputation in image processing pipelines}. In \bibinfo{booktitle}{\emph{Proceedings of the 34th ACM SIGPLAN Conference on Programming Language Design and Implementation}} (Seattle, Washington, USA) \emph{(\bibinfo{series}{PLDI '13})}. \bibinfo{publisher}{Association for Computing Machinery}, \bibinfo{address}{New York, NY, USA}, \bibinfo{pages}{519–530}.
\newblock
\showISBNx{9781450320146}
\urldef\tempurl%
\url{https://doi.org/10.1145/2491956.2462176}
\showDOI{\tempurl}


\bibitem[Roesch et~al\mbox{.}(2018)]%
        {roesch2018relay}
\bibfield{author}{\bibinfo{person}{Jared Roesch}, \bibinfo{person}{Steven Lyubomirsky}, \bibinfo{person}{Logan Weber}, \bibinfo{person}{Josh Pollock}, \bibinfo{person}{Marisa Kirisame}, \bibinfo{person}{Tianqi Chen}, {and} \bibinfo{person}{Zachary Tatlock}.} \bibinfo{year}{2018}\natexlab{}.
\newblock \showarticletitle{Relay: a new IR for machine learning frameworks}. In \bibinfo{booktitle}{\emph{Proceedings of the 2nd ACM SIGPLAN International Workshop on Machine Learning and Programming Languages}} (Philadelphia, PA, USA) \emph{(\bibinfo{series}{MAPL 2018})}. \bibinfo{publisher}{Association for Computing Machinery}, \bibinfo{address}{New York, NY, USA}, \bibinfo{pages}{58–68}.
\newblock
\showISBNx{9781450358347}
\urldef\tempurl%
\url{https://doi.org/10.1145/3211346.3211348}
\showDOI{\tempurl}


\bibitem[{S}amsung {A}dvanced {I}nstitute~of {T}echnology(2022a)]%
        {sait2022onemcc}
\bibfield{author}{\bibinfo{person}{{S}amsung {A}dvanced {I}nstitute~of {T}echnology}.} \bibinfo{year}{2022}\natexlab{a}.
\newblock \bibinfo{booktitle}{\emph{OneMCC}}.
\newblock
\urldef\tempurl%
\url{https://github.com/SAITPublic/OneMCC}
\showURL{%
\tempurl}


\bibitem[{S}amsung {A}dvanced {I}nstitute~of {T}echnology(2022b)]%
        {dev2023pimlibrary}
\bibfield{author}{\bibinfo{person}{{S}amsung {A}dvanced {I}nstitute~of {T}echnology}.} \bibinfo{year}{2022}\natexlab{b}.
\newblock \bibinfo{booktitle}{\emph{{PIMLibrary}}}.
\newblock
\urldef\tempurl%
\url{https://github.com/SAITPublic/PIMLibrary}
\showURL{%
\tempurl}


\bibitem[{S}amsung {A}dvanced {I}nstitute~of {T}echnology(2022c)]%
        {dev2022pimsimulator}
\bibfield{author}{\bibinfo{person}{{S}amsung {A}dvanced {I}nstitute~of {T}echnology}.} \bibinfo{year}{2022}\natexlab{c}.
\newblock \bibinfo{booktitle}{\emph{{PIMSimulator}}}.
\newblock
\urldef\tempurl%
\url{https://github.com/SAITPublic/PIMSimulator}
\showURL{%
\tempurl}


\bibitem[Shao et~al\mbox{.}(2022)]%
        {shao2022metaschedule}
\bibfield{author}{\bibinfo{person}{Junru Shao}, \bibinfo{person}{Xiyou Zhou}, \bibinfo{person}{Siyuan Feng}, \bibinfo{person}{Bohan Hou}, \bibinfo{person}{Ruihang Lai}, \bibinfo{person}{Hongyi Jin}, \bibinfo{person}{Wuwei Lin}, \bibinfo{person}{Masahiro Masuda}, \bibinfo{person}{Cody~Hao Yu}, {and} \bibinfo{person}{Tianqi Chen}.} \bibinfo{year}{2022}\natexlab{}.
\newblock \showarticletitle{Tensor Program Optimization with Probabilistic Programs}. In \bibinfo{booktitle}{\emph{Advances in Neural Information Processing Systems}}, \bibfield{editor}{\bibinfo{person}{S.~Koyejo}, \bibinfo{person}{S.~Mohamed}, \bibinfo{person}{A.~Agarwal}, \bibinfo{person}{D.~Belgrave}, \bibinfo{person}{K.~Cho}, {and} \bibinfo{person}{A.~Oh}} (Eds.), Vol.~\bibinfo{volume}{35}. \bibinfo{publisher}{Curran Associates, Inc.}, \bibinfo{pages}{35783--35796}.
\newblock
\urldef\tempurl%
\url{https://proceedings.neurips.cc/paper_files/paper/2022/file/e894eafae43e68b4c8dfdacf742bcbf3-Paper-Conference.pdf}
\showURL{%
\tempurl}


\bibitem[{S}hare{GPT} {T}eam(2023)]%
        {dev2023sharegpt}
\bibfield{author}{\bibinfo{person}{{S}hare{GPT} {T}eam}.} \bibinfo{year}{2023}\natexlab{}.
\newblock \bibinfo{booktitle}{\emph{ShareGPT}}.
\newblock
\urldef\tempurl%
\url{https://sharegpt.com, 2023}
\showURL{%
\tempurl}


\bibitem[Shen et~al\mbox{.}(2021)]%
        {shen2021nimble}
\bibfield{author}{\bibinfo{person}{Haichen Shen}, \bibinfo{person}{Jared Roesch}, \bibinfo{person}{Zhi Chen}, \bibinfo{person}{Wei Chen}, \bibinfo{person}{Yong Wu}, \bibinfo{person}{Mu Li}, \bibinfo{person}{Vin Sharma}, \bibinfo{person}{Zachary Tatlock}, {and} \bibinfo{person}{Yida Wang}.} \bibinfo{year}{2021}\natexlab{}.
\newblock \showarticletitle{Nimble: Efficiently Compiling Dynamic Neural Networks for Model Inference}. In \bibinfo{booktitle}{\emph{Proceedings of Machine Learning and Systems}}, \bibfield{editor}{\bibinfo{person}{A.~Smola}, \bibinfo{person}{A.~Dimakis}, {and} \bibinfo{person}{I.~Stoica}} (Eds.), Vol.~\bibinfo{volume}{3}. \bibinfo{pages}{208--222}.
\newblock
\urldef\tempurl%
\url{https://proceedings.mlsys.org/paper_files/paper/2021/file/5b47430e24a5a1f9fe21f0e8eb814131-Paper.pdf}
\showURL{%
\tempurl}


\bibitem[Shin et~al\mbox{.}(2023)]%
        {shin2023pimflow}
\bibfield{author}{\bibinfo{person}{Yongwon Shin}, \bibinfo{person}{Juseong Park}, \bibinfo{person}{Sungjun Cho}, {and} \bibinfo{person}{Hyojin Sung}.} \bibinfo{year}{2023}\natexlab{}.
\newblock \showarticletitle{PIMFlow: Compiler and Runtime Support for CNN Models on Processing-in-Memory DRAM}. In \bibinfo{booktitle}{\emph{Proceedings of the 21st ACM/IEEE International Symposium on Code Generation and Optimization}} (<conf-loc>, <city>Montr\'{e}al</city>, <state>QC</state>, <country>Canada</country>, </conf-loc>) \emph{(\bibinfo{series}{CGO 2023})}. \bibinfo{publisher}{Association for Computing Machinery}, \bibinfo{address}{New York, NY, USA}, \bibinfo{pages}{249–262}.
\newblock
\showISBNx{9798400701016}
\urldef\tempurl%
\url{https://doi.org/10.1145/3579990.3580009}
\showDOI{\tempurl}


\bibitem[SiFive(2021)]%
        {sifive}
\bibfield{author}{\bibinfo{person}{SiFive}.} \bibinfo{year}{2021}\natexlab{}.
\newblock \bibinfo{title}{SiFive E21 Core Complex Manual}.
\newblock
\newblock


\bibitem[Stone et~al\mbox{.}(2010)]%
        {stone2010opencl}
\bibfield{author}{\bibinfo{person}{John~E. Stone}, \bibinfo{person}{David Gohara}, {and} \bibinfo{person}{Guochun Shi}.} \bibinfo{year}{2010}\natexlab{}.
\newblock \showarticletitle{OpenCL: A Parallel Programming Standard for Heterogeneous Computing Systems}.
\newblock \bibinfo{journal}{\emph{Computing in Science \& Engineering}} \bibinfo{volume}{12}, \bibinfo{number}{3} (\bibinfo{year}{2010}), \bibinfo{pages}{66--73}.
\newblock
\urldef\tempurl%
\url{https://doi.org/10.1109/MCSE.2010.69}
\showDOI{\tempurl}


\bibitem[Taori et~al\mbox{.}(2023)]%
        {taori2023alpaca}
\bibfield{author}{\bibinfo{person}{Rohan Taori}, \bibinfo{person}{Ishaan Gulrajani}, \bibinfo{person}{Tianyi Zhang}, \bibinfo{person}{Yann Dubois}, \bibinfo{person}{Xuechen Li}, \bibinfo{person}{Carlos Guestrin}, \bibinfo{person}{Percy Liang}, {and} \bibinfo{person}{Tatsunori~B. Hashimoto}.} \bibinfo{year}{2023}\natexlab{}.
\newblock \bibinfo{booktitle}{\emph{Stanford Alpaca: An Instruction-following LLaMA model}}.
\newblock


\bibitem[Tillet et~al\mbox{.}(2019)]%
        {triton}
\bibfield{author}{\bibinfo{person}{Philippe Tillet}, \bibinfo{person}{H.~T. Kung}, {and} \bibinfo{person}{David Cox}.} \bibinfo{year}{2019}\natexlab{}.
\newblock \showarticletitle{Triton: an intermediate language and compiler for tiled neural network computations}. In \bibinfo{booktitle}{\emph{Proceedings of the 3rd ACM SIGPLAN International Workshop on Machine Learning and Programming Languages}} (Phoenix, AZ, USA) \emph{(\bibinfo{series}{MAPL 2019})}. \bibinfo{publisher}{Association for Computing Machinery}, \bibinfo{address}{New York, NY, USA}, \bibinfo{pages}{10–19}.
\newblock
\showISBNx{9781450367196}
\urldef\tempurl%
\url{https://doi.org/10.1145/3315508.3329973}
\showDOI{\tempurl}


\bibitem[UPMEM({[n.\,d.]})]%
        {upmemllvm}
\bibfield{author}{\bibinfo{person}{UPMEM}.} \bibinfo{year}{[n.\,d.]}\natexlab{}.
\newblock \bibinfo{title}{llvm-project}.
\newblock \bibinfo{howpublished}{\url{https://github.com/upmem/llvm-project}}.
\newblock


\bibitem[Vasilache et~al\mbox{.}(2018)]%
        {vasilache2018tc}
\bibfield{author}{\bibinfo{person}{Nicolas Vasilache}, \bibinfo{person}{Oleksandr Zinenko}, \bibinfo{person}{Theodoros Theodoridis}, \bibinfo{person}{Priya Goyal}, \bibinfo{person}{Zachary DeVito}, \bibinfo{person}{William~S. Moses}, \bibinfo{person}{Sven Verdoolaege}, \bibinfo{person}{Andrew Adams}, {and} \bibinfo{person}{Albert Cohen}.} \bibinfo{year}{2018}\natexlab{}.
\newblock \bibinfo{title}{Tensor Comprehensions: Framework-Agnostic High-Performance Machine Learning Abstractions}.
\newblock
\newblock
\showeprint[arxiv]{1802.04730}~[cs.PL]


\bibitem[Vogel et~al\mbox{.}(2018)]%
        {vogel2018fpga}
\bibfield{author}{\bibinfo{person}{Pirmin Vogel}, \bibinfo{person}{Andrea Marongiu}, {and} \bibinfo{person}{Luca Benini}.} \bibinfo{year}{2018}\natexlab{}.
\newblock \showarticletitle{Exploring shared virtual memory for FPGA accelerators with a configurable IOMMU}.
\newblock \bibinfo{journal}{\emph{IEEE Trans. Comput.}} \bibinfo{volume}{68}, \bibinfo{number}{4} (\bibinfo{year}{2018}), \bibinfo{pages}{510--525}.
\newblock


\bibitem[Wang(2021)]%
        {ben2021gptj}
\bibfield{author}{\bibinfo{person}{Ben Wang}.} \bibinfo{year}{2021}\natexlab{}.
\newblock \bibinfo{booktitle}{\emph{{Mesh-Transformer-JAX: Model-Parallel Implementation of Transformer Language Model with JAX}}}.
\newblock


\bibitem[XILINX(2020)]%
        {versalacap}
\bibfield{author}{\bibinfo{person}{XILINX}.} \bibinfo{year}{2020}\natexlab{}.
\newblock \bibinfo{title}{Versal ACAP AI Engine Architecture Manual}.
\newblock
\newblock


\bibitem[Yiu(2009)]%
        {yiu2009cortexm3}
\bibfield{author}{\bibinfo{person}{Joseph Yiu}.} \bibinfo{year}{2009}\natexlab{}.
\newblock \bibinfo{booktitle}{\emph{The definitive guide to the ARM Cortex-M3}}.
\newblock \bibinfo{publisher}{Newnes}.
\newblock


\bibitem[Zheng et~al\mbox{.}(2022)]%
        {dietcode}
\bibfield{author}{\bibinfo{person}{Bojian Zheng}, \bibinfo{person}{Ziheng Jiang}, \bibinfo{person}{Cody~Hao Yu}, \bibinfo{person}{Haichen Shen}, \bibinfo{person}{Josh Fromm}, \bibinfo{person}{Yizhi Liu}, \bibinfo{person}{Yida Wang}, \bibinfo{person}{Luis Ceze}, \bibinfo{person}{Tianqi Chen}, {and} \bibinfo{person}{Gennady Pekhimenko}.} \bibinfo{year}{2022}\natexlab{}.
\newblock \showarticletitle{DietCode: Automatic optimization for dynamic tensor program}.
\newblock  (\bibinfo{year}{2022}).
\newblock


\bibitem[Zheng et~al\mbox{.}(2020)]%
        {ansor2020zheng}
\bibfield{author}{\bibinfo{person}{Lianmin Zheng}, \bibinfo{person}{Chengfan Jia}, \bibinfo{person}{Minmin Sun}, \bibinfo{person}{Zhao Wu}, \bibinfo{person}{Cody~Hao Yu}, \bibinfo{person}{Ameer Haj-Ali}, \bibinfo{person}{Yida Wang}, \bibinfo{person}{Jun Yang}, \bibinfo{person}{Danyang Zhuo}, \bibinfo{person}{Koushik Sen}, \bibinfo{person}{Joseph~E. Gonzalez}, {and} \bibinfo{person}{Ion Stoica}.} \bibinfo{year}{2020}\natexlab{}.
\newblock \showarticletitle{Ansor: Generating {High-Performance} Tensor Programs for Deep Learning}. In \bibinfo{booktitle}{\emph{14th USENIX Symposium on Operating Systems Design and Implementation (OSDI 20)}}. \bibinfo{publisher}{USENIX Association}, \bibinfo{pages}{863--879}.
\newblock
\showISBNx{978-1-939133-19-9}
\urldef\tempurl%
\url{https://www.usenix.org/conference/osdi20/presentation/zheng}
\showURL{%
\tempurl}


\end{thebibliography}

\appendix
\section{Artifact Appendix}

\subsection{Abstract}
Our artifact includes setup and evaluation environments for the five configurations in Section~\ref{sec:methodology}: {\it CPU-autotuned}, {\it PrIM/(E)}, {\it PrIM+search}, {\it SimplePIM}, and {\it ATiM}, along with the full implementation of ATiM, including source code.
Scripts for configuration setup, autotuning, and evaluation are provided for each configuration. 
These scripts can be used to reproduce the main results presented in the paper (Figs.~\ref{fig:polybench_results}, \ref{fig:gpt_results}, and \ref{fig:opt_experiment}).
To enable quick reproduction of results, we provide pre-autotuned ATiM modules for tensor programs.
Users can also generate new ATiM modules by following the instructions provided below.
The top-level scripts are implemented by composing low-level scripts for an extensible and customizable evaluation workflow (see Section ~\ref{subsec:experiment_workflow}).  
Reproducing the experiments in Section~\ref{sec:evaluation} requires access to a UPMEM server.

\subsection{Artifact check-list (meta-information)}

{\small
\begin{itemize}[leftmargin=*,parsep=0in,topsep=0.05in]
  \item {\bf Algorithm: } Includes methods for (1) autotuning using ATiM-extended sketch generation rules and evolutionary search, (2) generating host and DPU code, including data transfer code, and (3) performing PIM-aware optimizations.
  \item {\bf Compilation: } Uses the \texttt{dpu-upmem-dpurte-clang} compiler from UPMEM SDK version 2021.3.0 with \texttt{-O2} optimization, LLVM 16.0.6 for TVM compilation and CPU code generation, and a modified Apache TVM compiler (based on version 0.13.0, commit 97c5de6) for ATiM.
  \item {\bf Model: } MTV and MMTV layers in the GPT-J.
  \item {\bf Run-time environment: } Linux (Ubuntu 20.04) with UPMEM driver, backends (communication library), and DPU runtime included in the UPMEM SDK version 2021.3.0.
  \item {\bf Hardware: } Intel Xeon Gold 5220R CPU server equipped with 20 DDR4-2400 PIM modules. The server requires UPMEM firmware (BIOS and MCU) provided by the UPMEM SDK.
  \item {\bf Metrics: } Execution time.
  \item {\bf Output: } \sloppy{Optimal tunable parameters for CPU-autotuned, PrIM/(E), PrIM+search, and SimplePIM and ATiM modules. Experimental results are stored as CSV files, and our scripts generate corresponding graphs in PDF format.}
  \item {\bf Experiments: } Detailed experimental procedures are described in Section~\ref{subsec:experiment_workflow} and the provided \texttt{README.md} file.
  \item {\bf How much disk space required (approximately)?: } \SI{10}{GB}.
  \item {\bf How much time is needed to prepare workflow (approximately)?: } 2 days to prepare tuned binaries and modules for PrIM/(E), PrIM+search, SimplePIM, CPU autotuning, and ATiM.
  \item {\bf How much time is needed to complete experiments (approximately)?: } 30 minutes.
  \item {\bf Publicly available?: } Yes (\url{https://github.com/SNU-CODElab/atim}).
  \item {\bf Workflow automation framework used?: } No.
  \item {\bf Archived (provide DOI)?: } \url{https://doi.org/10.5281/zenodo.15379924}.
\end{itemize}
}

\subsection{Description}

\subsubsection{How to access. }
Our source code and experimental scripts are available on Zenodo: \url{https://doi.org/10.5281/zenodo.15379924}. A GitHub repository is also provided (\url{https://github.com/SNU-CODElab/atim}).

\subsubsection{Hardware dependencies. }
We recommend using UPMEM systems with at least 16 PIM modules (2048 DPUs) to ensure sufficient inter-DPU parallelism. Our experiments use DDR4-2400 PIM modules, which are currently the only available UPMEM PIM modules.

\subsubsection{Software dependencies. }
We implemented and tested our codes on the Ubuntu 20.04 x86-64 system with UPMEM SDK version 2021.3.0.
Additional software dependencies include LLVM and TVM prerequisites on Ubuntu.
Specific software versions include Python 3.11, CMake 3.24 or higher, NumPy 1.26.4, XGBoost 1.6.1, and LLVM 16 (including clang).
We strongly recommend using the Docker image \texttt{yongwonshin/atim:v0.1} or later, as it includes all software dependencies and prerequisites required for running experiments.

\subsection{Installation}
\subsubsection{Using a Docker image. }
The \textit{docker} package is required. Activate the Docker image and install ATiM with the following commands:
\begin{lstlisting}[style=customminted, language=bash]
docker run -it --privileged yongwonshin/atim:v0.1
./install.sh
\end{lstlisting}

\subsubsection{Local Installation. }
ATiM source codes including PrIM and SimplePIM can be obtained via Zenodo or GitHub.

\begin{lstlisting}[style=customminted, language=bash]
# Extract from Zenodo:
wget -O atim.zip https://zenodo.org/records/15379924/files/atim.zip?download=1
unzip atim.zip

# Or clone from GitHub:
git clone -b artifact https://github.com/SNU-CODElab/atim.git
\end{lstlisting}

Detailed installation instructions are available in the \texttt{README.md} file in each archive.

\subsection{Experiment workflow }
\label{subsec:experiment_workflow}
The experiment workflow consists of executing parameter searches or autotuning processes for optimal tunable parameters and using them to compile tensor programs for each configuration.
Users may skip this step and directly proceed to Section~\ref{subsec:evaluation_and_expected_results} by reusing pre-tuned parameters and modules included in the artifact archive.
Users can follow the steps below to perform autotuning for CPU baseline and ATiM and parameter searches for PrIM/(E), PrIM+search, and SimplePIM:

\begin{lstlisting}[style=customminted, language=bash]
# Step 0: prepare for autotuning
cd atim/evaluation
conda activate atim-venv

# Step 1: perform autotuning for CPU-autotune
python cpu_autotune.py

# Step 2: find optimal parameters for PrIM
python prim_autotune.py

# Step 3: find optimal parameters for PrIM+search
python prim_search_autotune.py

# Step 4: find optimal parameter for SimplePIM
python simplepim_autotune.py

# Step 5: perform autotuning for ATiM
python atim_autotune.py
\end{lstlisting}

\subsection{Evaluation and expected results }
\label{subsec:evaluation_and_expected_results}
Once autotuning results are available, users can evaluate the execution times of tensor programs optimized using CPU-autotuned, PrIM/(E), PrIM+search, SimplePIM, and ATiM configurations.
Users can also evaluate the impact of different levels of ATiM's PIM-aware optimizations as defined in Section~\ref{subsection:upmem_effect_pim_aware_optim}. 

\begin{lstlisting}[style=customminted, language=bash]
# Evaluate tuned binaries/modules for tensor programs
python cpu_eval.py # CPU-autotuned
python prim_eval.py # PrIM/(E) and PrIM+search
python simplepim_eval.py # SimplePIM

python atim_eval.py # ATiM

# Evaluate ATiM's PIM-aware optimizations
python atim_branch_opt.py
\end{lstlisting}

\sloppy{Finally, the graphs in Figs.~\ref{fig:polybench_results}, \ref{fig:gpt_results}, and \ref{fig:opt_experiment} can be regenerated directly from the CSV files using the following command. The resulting graphs will be located in the \texttt{atim/evaluation/reproduced} directory.}

\begin{lstlisting}[style=customminted, language=bash]
python plot.py
\end{lstlisting}

\subsection{Experiment customization }

ATiM supports autotuning tensor programs across various workloads and tensor shapes. 
Users can run experiments on a specific workload with a custom tensor shape by supplying command-line arguments. For example, the following commands run autotuning and evaluation for an MMTV operation with a tensor shape of 256$\times$512$\times$256.

\begin{lstlisting}[style=customminted, language=bash]
python atim_autotune.py \
    --workload=mmtv --m=256 --n=512 --k=256

python atim_eval.py \
    --workload=mmtv --m=256 --n=512 --k=256
\end{lstlisting}
Additionally, users can modify the following files to test different workload sizes or introduce new workloads.

\begin{itemize}[leftmargin=*,parsep=0in,topsep=0.05in]
  \item \texttt{evaluation/bench.py}: Define new workloads in TIR.
  \item \texttt{evaluation/tasks.py}: Configure workload sizes.
  \item \texttt{evaluation/workloads.py}: Register workloads to run experiments.
\end{itemize}

\end{document}